\documentclass[11pt]{article}
\pdfoutput=1

\usepackage{jheppub}
\usepackage[utf8]{inputenc}
\usepackage{amsmath, bm, amssymb}
\usepackage{booktabs}
\usepackage{comment}
\usepackage{graphicx}
\usepackage{bbold}
\usepackage{subcaption}
\usepackage{mathtools}
\usepackage{cancel}
\usepackage[dvipsnames]{xcolor}
\usepackage{slashed}
\usepackage{multirow}
\usepackage{makecell}
\usepackage{tikz}
\usepackage{xspace}
\usepackage{numprint}
\usepackage[normalem]{ulem}
\usetikzlibrary{arrows.meta}
\usetikzlibrary{decorations.pathmorphing}

\numberwithin{equation}{section}

\newcommand\cF{\mathcal{F}}

\newcommand{\bK}{{\boldsymbol K}}
\newcommand\cO{\mathcal{O}}

\newcommand\beq{\begin{equation}}
\newcommand\eeq{\end{equation}}
\newcommand\beal{\begin{aligned}}
\newcommand\eeal{\end{aligned}}
\newcommand\bea{\begin{eqnarray}}
\newcommand\eea{\end{eqnarray}}

\newcommand{\nn}{\nonumber}

\newcommand{\vegas}{VEGAS\xspace}
\newcommand{\iflow}{\texttt{i-flow}\xspace}
\newcommand{\pysecdec}{\texttt{pySecDec}\xspace}
\newcommand{\fiesta}{\texttt{FIESTA}\xspace}

\newcommand\dd{{\mathrm d}}
\newcommand\cdotnew{\!\cdot\!}
\newcommand\vecbf[1]{{\boldsymbol #1}}

\definecolor{desycyan}{rgb}{0.00,0.68,0.93}
\definecolor{desyorange}{rgb}{0.93,0.58,0.16}

\makeatletter
\newcommand{\Biggg}{\bBigg@{3.5}}
\makeatother

\begin{document}
\preprint{
\texttt{DESY-22-144},
\texttt{IFT-UAM/CSIC-22-97},
\texttt{TUM-HEP\,1392/22}}
%\bigskip

\title{Machine Learning Post-Minkowskian Integrals}

\author[a,b]{Ryusuke Jinno,}
\author[b]{Gregor K\"alin,}
\author[b]{Zhengwen Liu,}
\author[b,c]{and Henrique Rubira}
\affiliation[a]{\small Instituto de F\'{\i}sica Te\'orica UAM/CSIC, \\
C/ Nicol\'as Cabrera 13-15, Campus de Cantoblanco, 28049 Madrid, Spain}
\affiliation[b]{\small Deutsches Elektronen-Synchrotron DESY, Notkestr.\,85, 22607 Hamburg, Germany}
\affiliation[c]{\small Physik Department T31, Technische Universit\"at M\"unchen,\\
James-Franck-Stra{\ss}e 1, D-85748 Garching, Germany}

\emailAdd{ryusuke.jinno@desy.de}
\emailAdd{gregor.kaelin@desy.de}
\emailAdd{zhengwen.liu@nbi.ku.dk}
\emailAdd{henrique.rubira@tum.de}

\abstract{%
We study a neural network framework for the numerical evaluation of Feynman loop integrals that
are fundamental building blocks for perturbative computations of physical observables in gauge and gravity theories.
We show that such a machine learning approach improves the convergence of the Monte Carlo algorithm for high-precision evaluation of multi-dimensional integrals compared to traditional algorithms.
In particular, we use a neural network to improve the importance sampling.
For a set of representative integrals appearing in the computation of the conservative dynamics for a compact binary system in General Relativity, we perform a quantitative comparison between the Monte Carlo integrators \vegas and \iflow, an integrator based on neural network sampling.
}
%\today~\currenttime

\maketitle

%\newpage
%%%%%%%%%%%%%%%%%%%%%%%%%%%%%%%%%%%%%%%%%%%%%%%%%%
\section{Introduction}\label{sec:intro}
%%%%%%%%%%%%%%%%%%%%%%%%%%%%%%%%%%%%%%%%%%%%%%%%%%

The success of gravitational-wave detections in the last decade \cite{LIGOScientific:2016aoc,LIGOScientific:2018mvr,LIGOScientific:2020ibl,LIGOScientific:2021djp} relies on our ability to construct high-precision waveform templates.
The most common gravitational wave sources are the binary inspiralling systems of black holes or/and Neutron stars.
Whereas we have seen exciting progress in numerical simulations, mostly for the merger phase, of binary systems \cite{Ajith:2012az,Szilagyi:2015rwa,Dietrich:2018phi} we discuss here a different type of numerical methods to compute certain constant ingredients for analytic approaches describing the binaries' movement.

Traditional approaches, performing a large-distance, small-velocity Post-Newtonian (PN) expansion (see e.g.\,\cite{Blanchet:2013haa,Schafer:2018kuf} for reviews), have been continuously pushing the state-of-the-art since the formulation of General Relativity a century ago.
The analytic output of these methods, describing e.g.\,the conservative motion of the constituents to high accuracy, is an essential input for the construction of waveforms.
More recently, constructions based on a worldline Effective Field Theory (EFT) formalism established by Goldberger and Rothstein \cite{Goldberger:2004jt} have started to compete with these traditional approaches \cite{Goldberger:2007hy,Foffa:2013qca,Rothstein:2014sra,Porto:2016pyg}.
This progress has resulted in the full knowledge of the conservative dynamics of non-spinning binary systems at the fourth perturbative order (4PN) from independent derivations in both approaches \cite{Foffa:2012rn,Damour:2014jta,Jaranowski:2015lha,Galley:2015kus,Bernard:2015njp,Porto:2017dgs,Porto:2017shd,Bernard:2017bvn,Marchand:2017pir,Foffa:2019yfl}.
Partial results at 5PN \cite{Foffa:2019hrb,Blumlein:2019zku,Foffa:2019eeb,Bini:2019nra,Blumlein:2020pyo,Blumlein:2021txe,Almeida:2021xwn} and 6PN \cite{Blumlein:2020znm,Bini:2020nsb,Bini:2020hmy,Blumlein:2021txj,Bini:2021gat} are also known.

Approaching the problem from a high-energy physicist's point of view lead to modern methods inspired by quantum field theory (QFT), reaching from worldline EFTs \cite{Kalin:2020mvi,Kalin:2020fhe,Kalin:2020lmz,Liu:2021zxr,Mogull:2020sak,Jakobsen:2021smu,Mougiakakos:2021ckm,Dlapa:2021npj,Jakobsen:2021lvp,Riva:2021vnj,Jakobsen:2021zvh,Jakobsen:2022fcj,Dlapa:2021vgp,Jakobsen:2022psy,Kalin:2022hph} to scattering-amplitude-based methods \cite{Neill:2013wsa,Vaidya:2014kza,Cheung:2018wkq,Bern:2019nnu,Bern:2019crd,Guevara:2018wpp,Kosower:2018adc,Maybee:2019jus,Damour:2019lcq,Cristofoli:2019neg,Bjerrum-Bohr:2018xdl,Haddad:2020que,Aoude:2020onz,Parra-Martinez:2020dzs,Bern:2020buy,Cheung:2020sdj,Cristofoli:2020uzm,Kosmopoulos:2021zoq,Bern:2021dqo,Kreer:2021sdt,Herrmann:2021lqe,DiVecchia:2021ndb,DiVecchia:2021bdo,Cristofoli:2021vyo,Bautista:2021wfy,Bjerrum-Bohr:2021wwt,Vanhove:2021zel,Bjerrum-Bohr:2021vuf,Mougiakakos:2020laz,Bjerrum-Bohr:2021din,Bern:2021yeh,Brandhuber:2021eyq}.
All these methods have in common that they describe the binary problem in the scattering regime and the expansion parameter is the gravitational coupling strength, i.e.\,Newton's constant $G$.
This resummation of all order velocity corrections at a given order in $G$ is called a \emph{Post-Minkowskian} (PM) expansion.
The \emph{potential} contributions to the scattering angle at the fourth PM (4PM) order~\cite{Dlapa:2021npj,Bern:2021dqo} have been extended by conservative \emph{tail} effects by the two different approaches~\cite{Dlapa:2021vgp,Bern:2021yeh}.
Very recently, the complete knowledge of the gravitational dynamics in the scattering of non-spinning bodies at 4PM order, incorporating conservative and dissipative effects \cite{Dlapa:2022lmu}, has been achieved by a combination of the worldline EFT approach and modern field theory techniques \cite{Dlapa:2023hsl}.
The (analytically determined) integrals used in \cite{Dlapa:2021vgp} were cross-checked by numerical methods discussed here.
Results for the hyperbolic (scattering) version of the two-body problem can be analytically continued to the elliptic case via a so-called boundary-to-bound map~\cite{Kalin:2019rwq,Kalin:2019inp}.
This map includes not only local conservative effects but also radiative corrections~\cite{Cho:2021arx,Saketh:2021sri}.
This map has been successfully checked against state-of-the-art PN results for bound orbits in the overlapping expansion region.

Multi-loop integrals are at the core of QFT methodologies.
Therefore, developing efficient techniques to evaluate these integrals is crucially important to advance the precision frontier for PM gravity.
The goal of this work is to study a set of (cut) Feynman integrals appearing in such approaches.
We will call them henceforth \emph{Post-Minkowskian} integrals.
In \cite{Kalin:2020fhe,Dlapa:2021npj} the generic structure of integrals needed for the computation of the deflection angle at 3PM and 4PM orders was identified, which easily generalizes to any order.
The main technique to compute, or rather bootstrap, such integrals used in these papers is the method of differential equations~\cite{Kotikov:1991pm,Remiddi:1997ny}, which reduces the calculation to finding the solution of a coupled system of first-order differential equations in one variable.
Whereas solving the differential equations is an art by itself (see e.g.\,\cite{Dlapa:2023hsl} for integrals discussed here), in some cases the boundary conditions turn out to be surprisingly tricky as well.

One application of numerical integration methods is to cross-check analytic results.
We develop here a machine-learning based framework for the numerical evaluation of multi-loop integrals, which is targeted to lay the groundwork for applications beyond simple cross-checking.
One can imagine that analytical methods will eventually hit a wall.
Numerical methods will provide a natural path forward for high-precision computations, for example via a hybrid analytical-numerical pipeline to efficiently produce waveform templates.
Pushing in that direction, we apply this novel method to numerically evaluate boundary values of PM integrals, which are a part of the pipeline for results in gravitational wave physics.
In the future, one could try to directly determine boundary conditions to the differential equation system with numerical methods, inputting them either as high precision constants to the final answer or as a way to conjecture its analytical form via \emph{integer relation} algorithms like PSLQ~\cite{bailey1991polynomial,Bailey:1999nv}.
The latter strategy was for example successfully applied in a similar computation in~\cite{Bini:2020uiq}.

Due to the use of dimensional regularization -- meaning that we compute integrals in $D=4-2\epsilon$ dimensions -- such boundary integrals depend on $\epsilon$.
Since we are only interested in $\epsilon$-divergent and -finite contributions to an observable it is sufficient to compute the boundary integrals up to a certain order as a power series in $\epsilon$.
\emph{Sector decomposition}~\cite{Prokhorenko:2007yy,Roth:1996pd,Binoth:2000ps,Heinrich:2008si} is a method to perform such a power series expansion on an integrand level by breaking the integral into smaller pieces, so-called sectors.
Many tools like ({{\tt py}){\tt SecDec}~\cite{Carter:2010hi,Borowka:2012yc,Borowka:2015mxa,Borowka:2017idc} or {\tt FIESTA}~\cite{Smirnov:2008py,Smirnov:2009pb,Smirnov:2013eza,Smirnov:2015mct,Smirnov:2021rhf} implement sector decomposition methods together with numerical integration algorithms.
We used these programs to produce decomposed integrands, which we then integrated with machine learning techniques implemented in \iflow~\cite{Gao:2020vdv}.
The main idea of \iflow is to use a \emph{neural network} (NN) to improve the Monte-Carlo integration and (importance) sampling, which improves the error estimates and leads to faster convergence of the numerical integration.
\iflow uses the method of \emph{normalizing flows} \cite{dinh2015nice,muller2019neural}, which approximates the phase-space integrand via an (analytically) invertible neural network. 

The main result of this paper consists of a quantitative analysis of the required number of integrand evaluations to reach a given accuracy goal, comparing traditional sampling methods such as \vegas \cite{Lepage:1977sw, Lepage:2020tgj} to our neural-network-based framework.
We analyze a representative set of Post-Minkowskian boundary integrals (in the so-called \emph{potential} region) reaching from two (3PM) to four loops (5PM).
Since the neural network needs to be \emph{trained} for a constant initial time they perform worse for low relative precision ($\sim 10^{-3}$) but start to scale significantly better for higher precision ($\sim 10^{-4}$ and below).  

We begin in Sec.\,\ref{sec:theory} by introducing the loop families of interest and %present (partially novel)
list
analytical results for most boundary master integrals up to three loops, and a few representative four-loop integrals.
The sector decomposition methods and our numerical setup, mostly focused on machine learning techniques, are introduced in Sec.~\ref{sec:numerical}.
This section also contains the main results for our numerical integration framework.
In Sec.~\ref{sec:discussion} our findings are summarized and we conclude with a perspective into future applications of machine learning techniques to Feynman integration.

%%%%%%%%%%%%%%%%%%%%%%%%%%%%%%%%%%%%%%%%%%%%%%%%%%
\section{Post-Minkowskian integrals}\label{sec:theory}
%%%%%%%%%%%%%%%%%%%%%%%%%%%%%%%%%%%%%%%%%%%%%%%%%%
This section introduces a set of Feynman integrals appearing in field theory based approaches to gravitational binary dynamics.\footnote{Analytic derivations of many of the integrals presented here are also discussed in \cite{Dlapa:2023hsl}. We reproduce some of the derivations (and more) in this section and in the appendix for self-consistency reasons.}
We present a representative set of loop integrals and their analytic expressions.
For one, two, and three loops those correspond to master integrals with respect to integration-by-parts relations.
We will then apply machine learning techniques to numerically evaluate them in subsequent sections. 
We restrict ourselves to the first three orders in the $\epsilon$ series for numerical checks.

\subsection{Prerequisites}
At $\mathcal{O}(G^{L+1})$ order we define the set of \emph{Post-Minkowskian integrals} by \cite{Dlapa:2023hsl}
\begin{align}\label{PMint}
I^{(a_1 \cdots a_L; \pm\cdots\pm)}_{\alpha_1 \cdots \alpha_L; \nu_1 \cdots \nu_N}(\gamma) \,=\, 
\int\Bigg(\prod_{i=1}^{L} {\dd^D\!\ell_i\, \frac{e^{\epsilon \gamma_E}}{\pi^{(D-1)/2}} }\, \frac{\delta(\ell_i \cdotnew u_{a_i})}{(\pm \ell_i\cdotnew u_{\slashed{a}_i} \! - i0)^{\alpha_i}} \Bigg) 
\frac{(-q^2)^{\nu -L(D-1)/2}}{P_1^{\nu_1} P_2^{\nu_2} \cdots P_N^{\nu_N}},
\end{align}
where $\alpha_i, \nu_r \in\mathbb{Z}$, $\nu=(\alpha_1 + \cdots + \alpha_L)/2 + \nu_1 + \cdots + \nu_N$, $\ell_i$ stand for loop momenta, $a_i\in\{1,2\}$ and $\slashed{a}_i= a_i - (-1)^{a_i}$.
We adopt the mostly minus Minkowski metric, $\eta_{\mu\nu}=\operatorname{diag}(1,-1,-1,-1)$, and work in dimensional regularisation in $D= 4-2\epsilon$ dimensions.
We introduced a convenient normalization factor $e^{\epsilon \gamma_E}$ per loop, where $\gamma_E$ is the Euler–Mascheroni constant.
The inverse propagators $P_i$ (including irreducible scalar products for $\nu_i<0$) can be expressed in terms of the external and loop momenta
\begin{align}\label{}
P_i = -(\lambda_{ij} \ell_j + \beta_i q)^2 - i0, \quad \lambda_{ij}, \beta_i \in\{0,\pm 1\}, \quad
1\leqslant i \leqslant N= \frac{L(L {+} 3)}{2}.
\end{align}
We use implicit `$-i0$' prescriptions for all propagators in the rest of the paper.
The external kinematical variables satisfy
\begin{align}\label{kinematics}
q\cdot u_1 = q\cdot u_2=0,\qquad
u_1^2 = u_2^2 = 1.
\end{align}
A useful property is that there is a single dimensionful kinematical variable $t=-q^2 = \vecbf{q}^2$ in the integrals.
Thus, the dependence on $t$ can be easily fixed by the mass dimension and is given by $t^{L(D-1)/2-\nu}$.
As a result, the integrals in \eqref{PMint} are dimensionless functions of a single variable $\gamma=u_1\cdot u_2$, where in the scattering region $\gamma>1$.

An atypical feature of the integrals in \eqref{PMint} is that each loop integration is partially localized by a Dirac-delta constraint, whose argument is linear in the loop momentum and one of the initial velocities of the bodies $\delta(\ell_i\cdot u_a)$.
Similar loop integrals appear in PM methods relying on gravitational scattering amplitudes~\cite{Cheung:2018wkq,Bern:2019nnu,Bern:2019crd,Cheung:2020gyp,Cheung:2020sdj,Bern:2020buy,Kosmopoulos:2021zoq,Bern:2020uwk,Bern:2021dqo,Bern:2021yeh}.
They are related to the PM integrals in \eqref{PMint} by so-called `reverse unitarity' \cite{Cutkosky:1960sp, Anastasiou:2002yz, Anastasiou:2003ds}, in which a Dirac-delta function is understood as a cut of a propagator.
Thus, many techniques, including the novel numerical techniques developed in this work, are applicable for loop integrals in both worldline EFT and S-matrix-based formulations.

It was found that the method of differential equations~\cite{Kotikov:1991pm,Remiddi:1997ny} provides an efficient way to determine the $\gamma$-dependency of PM integrals~\cite{Parra-Martinez:2020dzs,Kalin:2020fhe,Dlapa:2021vgp,Dlapa:2023hsl}.
Using integration-by-parts (IBP) relations~\cite{Tkachov:1981wb,Chetyrkin:1981qh, Anastasiou:2004vj}, one can derive a system of \emph{ordinary differential equations} with respect to the kinematical variable $\gamma$ for a set of basis (master) integrals.
To be clear, let us take a look at the simplest example where the same velocity vector $u_a$ ($a=1$ or $a=2$) appears in all delta-function constraints in \eqref{PMint}.
In this case, any integral obeys the following simple differential equation:
\begin{align}\label{}
\frac{\dd}{\dd\gamma}
I^{(2 \cdots 2)}_{\alpha_1 \cdots \alpha_L; \nu_1 \cdots \nu_N} (\gamma)
\,=\,   \frac{-\gamma \sum_{j=1}^L\alpha_i}{\gamma^2 -1}\, I^{(2 \cdots 2)}_{\alpha_1 \cdots \alpha_L; \nu_1 \cdots \nu_N} (\gamma).
\end{align}
We can immediately write down its solution
\begin{align}\label{Schw-int-sol}
I^{(2 \cdots 2)}_{\alpha_1 \cdots \alpha_L; \nu_1 \cdots \nu_N}(\gamma) 
\,=  (\gamma^2 - 1)^{-\frac{1}{2}\sum_{j=1}^L \alpha_j}\,
\vecbf{I}_{\alpha_1 \cdots \alpha_L; \nu_1 \cdots \nu_N},
\end{align}
where $\vecbf{I}_{\alpha_1 \cdots \alpha_L; \nu_1 \cdots \nu_N}$ is the boundary value of $I_{\alpha_1 \cdots \alpha_L; \nu_1 \cdots \nu_N}$ in the static limit $\gamma\to 1$.
These boundary integral are defined in Euclidean space of $d=D-1$ dimensions
\begin{align}\label{Schw-int-static}
\vecbf{I}_{\alpha_1 \cdots \alpha_L; \nu_1 \cdots \nu_N} \,\equiv\
\int\Bigg(\prod_{j=1}^{L} {\dd^{d} \ell_j \over \pi^{d/2} }\,
{e^{\gamma_E \epsilon} \over  (\pm \ell_j^z - i0)^{\alpha_j}} \Bigg) 
{(\vecbf{q}^2)^{\nu - Ld/2} \over \vecbf{P}_1^{\nu_1} \vecbf{P}_2^{\nu_2} \cdots \vecbf{P}_N^{\nu_N}},
\end{align}
where $\vecbf{P}_i$ is the $d$-dimensional part of ${P}_i$, i.e.~the time component removed.
On one hand, these integrals contribute to the test-particle limit (geodesic motion in a Schwarzschild background for the spin-less case).
On the other hand, more interestingly, in the $\gamma\rightarrow 1$ potential region~\cite{Beneke:1997zp,Smirnov:1998vk,Smirnov:1999bza,Jantzen:2012mw} all integrals of the form \eqref{PMint} from other sectors can be reduced to \eqref{Schw-int-static} as well.
To be precise, if we are working in the rest frame of the particle 2,
\begin{align}\label{u-rest-frame}
u_1^\mu = \gamma (1, 0,0, \beta),~~  u_2^\mu = (1, 0,0,0)
~~~\text{with}~~\beta= \gamma^{-1}\sqrt{\gamma^2 - 1}
\end{align}
upon resolving the delta-function constraints $\delta(\ell_i \cdotnew u_1)\delta(\ell_j \cdotnew u_2)$ one finds $\ell_i^0=\beta\ell_i^z$ and $\ell_j^0=0$.
Therefore, using this frame and expanding the integrand around the small velocity limit $\beta\to 0$ or $\gamma\to 1$ leads to
\begin{align}\label{}
&
{1 \over \pm\ell_i\cdotnew u_2 - i0} = {1 \over \beta}\, {1 \over \pm\ell_i^z - i0}, \qquad
{1 \over \pm\ell_j\cdotnew u_1 - i0} = {1 \over \beta}\, {1 \over \mp\ell_j^z - i0},
\\[0.2 em]
&
{1 \over -(\ell_i {+} \ell_j {-} q)^2 - i0} = {1 \over -(\beta\ell_i^z)^2 + (\vecbf{\ell}_i {+} \vecbf{\ell}_j {-} \vecbf{q})^2 - i0}
= {1 \over (\vecbf{\ell}_i {+} \vecbf{\ell}_j {-} \vecbf{q})^2 - i0} + \mathcal{O}(\beta^2).
\label{soft-expansion}
\end{align}
We refer to the integrals defined in \eqref{Schw-int-static} as {\it static integrals}.
They play a crucial role in evaluating PM integrals in the context of the differential equation method as they encode all boundary data in the potential region.
They are the {\it core objects} of interest in this work.
We list a representative set of static integrals and their analytic results in the following subsections.

%**************************************************
\subsection{2PM: One loop}\label{sec:pm-int-1L}

At one-loop level, all static integrals can be immersed into the following form
\begin{align}\label{static-1-loop}
\vecbf{A}_{\alpha \nu_1 \nu_2} = e^{\epsilon\gamma_E}
\int {\dd^{d} \ell \over \pi^{d/2} }\,{(\vecbf{q}^2)^{\nu_1+\nu_2+\alpha/2-d/2} \over (\pm\ell^z)^{\alpha} (\vecbf{\ell}^2)^{\nu_1}\, [(\vecbf{\ell} {-} \vecbf{q})^2]^{\nu_2}}.
\end{align}
These integrals are sufficient for the computation of the conservative dynamics of non-spinning \cite{Kalin:2020mvi} and spinning \cite{Liu:2021zxr} binary systems at $\mathcal{O}(G^2)$.
Any integral in \eqref{static-1-loop} is independent of the sign in front of the linear propagator $\pm \ell^z-i0$, where we have written out the otherwise implicit $-i0$.

Via IBP relations any integral of the form \eqref{static-1-loop} can be expressed in terms of two master integrals $\{\vecbf{A}_{011}, \vecbf{A}_{111}\}$.
Technically, it is not necessary to perform any IBP reduction since the analytical expression for generic $\{\alpha, \nu_1,\nu_2\}$ ($\nu_1>0$, $\nu_2> 0$) and $d$ is known \cite{Smirnov:2012gma}
\begin{align}\label{pm-1-loop}
\vecbf{A}_{\alpha \nu_1 \nu_2} &=  e^{\gamma_E\epsilon}\,
{2^{\alpha -1} i^{\alpha}\, \Gamma(\alpha/2)\, \Gamma({d - \alpha \over 2} - \nu_1)\,
 \Gamma({d-\alpha \over 2} - \nu_2)\, \Gamma({\alpha - d \over 2} +\nu_1+\nu_2)
\over \Gamma(\alpha)\, \Gamma(\nu_1)\,\Gamma(\nu_2)\,\Gamma (d-\alpha-\nu_1-\nu_2)}.
\end{align}
We have merely presented this result for completeness and we are not interested in their numerical evaluation.
%In later section we will also use this result in a recursive way to find analytic expressions for higher-loop integrals.
%Important in that aspect is that this result only holds for the special kinematics we are considering here, i.e.\,$\vecbf{q}\cdot\vecbf{u}=0$ with $\vecbf{u}$ the unit vector perpendicular to $\vecbf{q}$ which in our chosen frame \eqref{u-rest-frame} reduces to $\vecbf{q}\cdot\vecbf{u}=q^z=0$, and can, thus, not in all cases be used in a recursive loop-by-loop fashion.

\subsection{3PM: Two loops}\label{sec-pm-int-2L}

At two-loop order, all static integrals can be mapped into the following family \cite{Kalin:2020fhe,Kalin:2020lmz}
\begin{align}\label{Kite-3d}
\vecbf{K}&^{(\pm\pm)}_{\alpha_1 \alpha_2; \nu_1 \cdots \nu_5}
\\
&= 
\int {\dd^{d} \ell_1 \dd^{d} \ell_2 \over \pi^{d}}\, 
{e^{2\epsilon\gamma_E}\, (\vecbf{q}^2)^{\nu_1 + \cdots +\nu_5 + (\alpha_1 {+} \alpha_2)/2 - d} \over 
(\pm \ell_1^z)^{\alpha_1} (\pm \ell_2^z)^{\alpha_2}\,
[\vecbf{\ell}_1^2]^{\nu_1}
[\vecbf{\ell}_2^2]^{\nu_2}
[(\vecbf{\ell}_{12} {-} \vecbf{q})^{2}]^{\nu_3}
[(\vecbf{\ell}_1 {-} \vecbf{q})^2]^{\nu_4}
[(\vecbf{\ell}_2 {-} \vecbf{q})^2]^{\nu_5}
},
\nonumber
\end{align}
where we denote $\ell_{i\cdots j} = \ell_i + \cdots + \ell_j$.
The five squared propagators in \eqref{Kite-3d} graphically correspond to the Kite topology:
\begin{align*}\label{}
\begin{aligned}
\begin{tikzpicture}[scale=0.8]
  \draw[line width=0.8pt] (-2,0) -- (0, 1.0) -- (2,0) -- (0,-1.0) -- (-2, 0); 
  \draw[line width=0.8pt] (0, 1.0) -- (0, -1.0);
  \draw[line width=1.2pt] (-1.99,0) -- (-3,0);
  \draw[line width=1.2pt] (1.99,0) -- (3,0);
\end{tikzpicture}\,.
\end{aligned}
\end{align*}

Solving IBP identities using \texttt{FIRE6}/\texttt{LiteRed} \cite{Smirnov:2019qkx,Lee:2012cn,Lee:2013mka} or \texttt{Kira2} \cite{Klappert:2020nbg}, we find that 9 independent master integrals for all sign configurations of linear propagators in \eqref{Kite-3d}.
As expected, each master integral has a either double-bubble or sunrise topology when considering only square-type propagators:
\begin{align*}\label{}
\begin{aligned}
\begin{tikzpicture}[scale=0.8]
  \newcommand{\nicearrowx}{-{Latex[length=10mm, width=2mm]}}
  \draw[xshift=-24pt, line width=0.8pt] (0,0) ellipse (24pt and 20pt); 
  \draw[xshift=24pt, line width=0.8pt] (0,0) ellipse (24pt and 20pt); 
  \draw[xshift=-48pt, -, line width=1.2pt] (0,0) -- (-1.2,0);
  \draw[xshift=48pt, -, line width=1.2pt] (0,0) -- (1.2,0);
  \draw[xshift=170pt, line width=0.8pt] (0,0) circle (8 mm);  
  \draw[xshift=170pt, -, line width=0.8pt] (-1.8,0) -- (1.8,0);
  \draw[xshift=170pt, -, line width=1.2pt] (-1.8,0) -- (-0.8,0);
  \draw[xshift=170pt, -, line width=1.2pt] (1.8,0) -- (0.8,0);
\end{tikzpicture}\,.
\end{aligned}
\end{align*}
We list all their analytical results below:
\begingroup
\allowdisplaybreaks
\begin{align}
%1
\bK_{00;00111} &= e^{2\epsilon \gamma_\textrm{E}}\frac{\Gamma^3(1/2-\epsilon)\, \Gamma(2\epsilon)}{\Gamma(3/2-3\epsilon)} 
= \frac{\pi }{\epsilon}+6 \pi - \pi\left(\frac{7}{6}\pi^2 - 36\right) \epsilon +\cO (\epsilon^2)\,,
\label{eq:K2_1}
\\[0.7 em]
%2
\bK_{00;11011} &= e^{2\epsilon \gamma_\textrm{E}}\frac{\Gamma^4(1/2-\epsilon)\,\Gamma^2(1/2+\epsilon)}{\Gamma^2(1-2\epsilon)}
\label{eq:K2_2}
\\
&=\pi^3+4 \pi^3 \epsilon \log (2)+ \pi^3\left(\frac{5 \pi^2}{6}+8  \log^2(2)\right)\epsilon^2 +\cO (\epsilon^3)\,,
\nn\\[0.7 em]
%3
\bK^{(\pm)}_{01;00111} &= i\sqrt{\pi}e^{2\epsilon\gamma_\textrm{E}}\frac{\Gamma(1/2-2\epsilon)\,\Gamma^2(1/2-\epsilon)\,\Gamma(-\epsilon)\,\Gamma(1/2+2\epsilon)}{\Gamma(1/2-3\epsilon)\,\Gamma(1-2\epsilon)}
\label{eq:K2_3}
\\
&=-i \pi^2 \bigg[\frac{1}{\epsilon}- 2 \log (2) + 2 \epsilon \log^2(2) \bigg]+\cO (\epsilon^2)\,,
\nn\\[0.7 em]
%4
\bK^{(\pm)}_{01;11011} &= i\sqrt{\pi}e^{2\epsilon\gamma_\textrm{E}}\frac{\Gamma^2(1/2-\epsilon)\,\Gamma^2(-\epsilon)\,\Gamma(1/2+\epsilon)\,\Gamma(1+\epsilon)}{\Gamma(1-2\epsilon)\,\Gamma(-2\epsilon)}
\label{eq:K2_4}
\\
&=-2 i \pi^2 \bigg[\frac{1}{\epsilon} + 2\log (2) + \frac{1}{3} \epsilon \left(\pi^2 + 6 \log^2(2)\right) \bigg]+\cO (\epsilon^2)\,,
\nn
\\[0.7 em]
%5
\bK^{(\pm)}_{01;10110} &= i\pi 2^{6\epsilon}e^{2\epsilon\gamma_\textrm{E}}\frac{\Gamma(\epsilon)\,\Gamma(1/2-2\epsilon)\,\Gamma(1/2+2\epsilon)}{\Gamma(1-\epsilon)}
\label{eq:K2_5}
\\
&= i \pi^2\bigg[ \frac{1}{\epsilon}+6  \log (2)+2  \epsilon \left(\pi^2+9 \log^2(2)\right) \bigg]+\cO (\epsilon^2)\,,
\nn\\[0.7 em]
%6
\bK^{(+-)}_{11;00111} &= -e^{2\epsilon\gamma_\textrm{E}}\frac{2\pi}{3}\frac{\Gamma^3(-\epsilon)\,\Gamma(2\epsilon+1)}{\Gamma(-3\epsilon)}
= -\frac{2 \pi }{\epsilon^2}+\frac{\pi^3}{3}+\cO (\epsilon)\,,
\label{eq:K2_6}
\\[0.7 em]
%7
\bK^{(++)}_{11;00111} &= 2 \bK^{(+-)}_{11;00111}
= -\frac{4 \pi }{\epsilon^2}+\frac{2 \pi^3}{3}+\cO (\epsilon)\,,
\label{eq:K2_7}
\\[0.7 em]
%8
\bK^{(++)}_{11;11011} &= - e^{2\epsilon\gamma_\textrm{E}} \frac{\pi\,\Gamma^4(-\epsilon)\,\Gamma^2(\epsilon+1)}{\Gamma^2(-2\epsilon)}
= -\frac{4 \pi }{\epsilon^2}+\frac{2 \pi^3}{3}+\cO (\epsilon)\,,
\label{eq:K2_8}
\\[0.7 em]
%9
\bK^{(\pm)}_{02;10110} &= - e^{2\epsilon\gamma_\textrm{E}}\frac{4\epsilon\, \Gamma(2\epsilon)\,\Gamma^2(-2\epsilon)\, \Gamma(1/2-\epsilon)\,\Gamma(1/2+\epsilon)}{\Gamma(-4\epsilon)}
=\frac{2 \pi }{\epsilon}+\frac{\pi^3 \epsilon}{3}+\cO (\epsilon^2)\,,
\label{eq:K2_9}
\end{align}
\endgroup
where the sign superscript is omitted in case a linear propagator is not present. 
These results were used in~\cite{Kalin:2020fhe,Kalin:2020lmz} and an analytical derivation is presented in~\cite{Dlapa:2023hsl}.
Most of them can be computed by using the one-loop formula \eqref{static-1-loop} iteratively loop-by-loop, including \eqref{eq:K2_1}, \eqref{eq:K2_2}, \eqref{eq:K2_3}, \eqref{eq:K2_4} and \eqref{eq:K2_8}.
Integrals \eqref{eq:K2_5} and \eqref{eq:K2_9} can be similarly obtained via a loop-by-loop integration.
Computing $\bK^{+\pm}_{11;00111}$ in \eqref{eq:K2_6} and \eqref{eq:K2_7} is not as trivial.
Two independent derivations -- one based on a symmetrization trick and one via direct integration of a Feynman parametrized form -- are presented in App.\,\ref{sec:appA}.
The latter rather considers a generalized version of this integral with generic symbolic indices for some slots. 
The resulting expression needs some non-trivial transformation in order to lead to the simple form presented here.

%**************************************************
\subsection{4PM: Three loops}\label{sec:pm-int-3L}
At three-loop level, all static integrals appearing in the computation of the next-to-next-to-next-to-leading order conservative dynamics of non-spinning binaries \cite{Dlapa:2021npj, Dlapa:2021vgp} can be reduced to the following three topologies (of squared propagators) \cite{Dlapa:2023hsl}:
\begin{align}\label{static-topo-3L}
\begin{aligned}
\begin{tikzpicture}[scale=0.7] 
  \draw[line width=0.8pt] (0,0) arc (10:170:2 and 1.2);
  \draw[line width=0.8pt] (0,0) arc (10:170:2 and 0.4);
  \draw[line width=0.8pt] (0,0) arc (-10:-170:2 and 0.4);
  \draw[line width=0.8pt] (0,0) arc (-10:-170:2 and 1.2);
  \draw[xshift=0cm, -, line width=1.2pt] (0,0) -- (0.8,0);
  \draw[xshift=-39.2mm, -, line width=1.2pt] (0,0) -- (-0.8,0); 
  \fill[yshift=-15mm] (-2,0) circle (0) node[] {$(\vecbf{B})$};
%%%%
%%%%  
  \draw[xshift=47mm, xshift=10mm, line width=0.8pt] (0,0) circle (1.0);
  \draw[xshift=47mm, xshift=-10mm, line width=0.8pt](0,0) circle (1.0);
  \draw[xshift=47mm, -, line width=0.8pt] (0,0) -- (2.0,0);  
  \draw[xshift=47mm, xshift=20mm, -, line width=1.2pt] (0,0) -- (0.8,0);
  \draw[xshift=47mm, xshift=-20mm, -, line width=1.2pt] (0,0) -- (-0.8,0); 
  \fill[xshift=47mm, yshift=-15mm] (0,0) circle (0) node[] {$(\vecbf{C})$};
%%%%
%%%%
  \draw[xshift=113mm, xshift=-10mm, line width=0.8pt] (0,0) ellipse (1.0 and 0.5);
  \draw[xshift=113mm, xshift=10mm, line width=0.8pt] (0,0) ellipse (1.0 and 0.5);
  \draw[xshift=113mm, xshift=20mm, line width=0.8pt] (0,0) arc (-2:182:2 and 1.2);
  \draw[xshift=113mm, xshift=20mm, -, line width=1.2pt] (0,0) -- (0.8,0);
  \draw[xshift=113mm, xshift=-20mm, -, line width=1.2pt] (0,0) -- (-0.8,0); 
  \fill[xshift=113mm, yshift=-15mm] (0,0) circle (0) node[] {$(\vecbf{D})$};
\end{tikzpicture}\,.
\end{aligned}
\end{align}

In the following we will denote integrals by their topology, a subscript counter, and the usual superscript of signs of linear propagators.
Let us start with integrals without linear propagators:
\begin{align} 
\vecbf{B}_{0}  &= e^{3\epsilon\gamma_E}\,
\int{\dd^d\ell_1 \dd^d\ell_2 \dd^d\ell_3 \over \pi^{3d/2}}\,
{(\vecbf{q}^2)^{4 - 3d/2} \over \vecbf{\ell}_1^2\, \vecbf{\ell}_2^2\, \vecbf{\ell}_3^2\, (\vecbf{\ell}_{123} {-} \vecbf{q})^2}, \label{eq:3L_B}
\\[0.5 em]
\vecbf{C}_{0}  &= e^{3\epsilon\gamma_E}\,
\int{\dd^d\ell_1 \dd^d\ell_2 \dd^d\ell_3 \over \pi^{3d/2}}\,
{(\vecbf{q}^2)^{5 - 3d/2} \over \vecbf{\ell}_1^2\, \vecbf{\ell}_2^2\, \vecbf{\ell}_3^2\, (\vecbf{\ell}_1 {-} \vecbf{q})^2\, (\vecbf{\ell}_{23} {-} \vecbf{q})^2}, \label{eq:3L_C}
\\[0.5 em]
\vecbf{D}_{0} &= e^{3\epsilon\gamma_E}\,
\int{\dd^d\ell_1 \dd^d\ell_2 \dd^d\ell_3 \over \pi^{3d/2}}\,
{(\vecbf{q}^2)^{5- 3d/2} \over \vecbf{\ell}_1^2\, \vecbf{\ell}_2^2\, \vecbf{\ell}_3^2\, (\vecbf{\ell}_{13} {-} \vecbf{q})^2\, 
(\vecbf{\ell}_{23} {-} \vecbf{q})^2}.\label{eq:3L_D}
\end{align}
In this case, it is clear that they can be evaluated to a product of Gamma functions using the one-loop bubble \eqref{pm-1-loop} iteratively.
They explicitly evaluate to:
\begin{align}\label{}
\vecbf{B}_{0}  &= e^{3\epsilon\gamma_E}\,
\frac{\Gamma^4({1/2}-\epsilon)\, \Gamma(3 \epsilon - {1/2})}{\Gamma (2-4 \epsilon)},
\nonumber\\
&= -2 \pi^{5/2}
\bigg[1 + 2\epsilon(5+\log2)
+
\epsilon ^2\left(76+\frac{23\pi^2}{12} + 2(10+\log2)\right)\log2
\Bigg]
+ \mathcal{O}(\epsilon^3),
\\[0.5 em]
\vecbf{C}_{0}  &= e^{3\epsilon\gamma_E}\,
{\Gamma (2\epsilon)\, \Gamma(\epsilon + {1/2})\, \Gamma^5({1/2}-\epsilon) \over \Gamma({3/2} - 3\epsilon)\, \Gamma(1-2\epsilon)}
\nonumber\\
&= \pi^{5/2} \bigg[
\frac{1}{\epsilon} +2(3+\log2)
+\epsilon \left(36-\frac{3 \pi ^2}{4}+2(6+\log 2)\log 2\right)
\bigg]
+ \mathcal{O}(\epsilon^2),
\\[0.5 em]
\vecbf{D}_{0} &= e^{3\epsilon\gamma_E}\,
{\Gamma({1/2} - 3\epsilon)\,\Gamma({1/2} + 3\epsilon)\, \Gamma^5({1/2} - \epsilon)\, \Gamma^2({1/2} + \epsilon)
\over  \Gamma(1-4\epsilon)\, \Gamma^2(1-2\epsilon)\, \Gamma(1 + 2\epsilon)}
\nonumber\\
&= \pi^{9/2}\bigg[1 + 6\epsilon\log2
+ \left(\frac{47 \pi ^2}{12}+18 \log^22\right)\epsilon^2
\bigg]
+ \mathcal{O}(\epsilon^3).
\end{align}
For the case of one linear propagator, we find the following master integrals:
\begin{align}\label{}
\vecbf{B}_1  &= e^{3\epsilon\gamma_E}\,
\int{\dd^d\ell_1 \dd^d\ell_2 \dd^d\ell_3 \over \pi^{3d/2}}\,
{(\vecbf{q}^2)^{4 + (1- 3d)/2} \over (\pm\ell_3^z)\, \vecbf{\ell}_1^2\, \vecbf{\ell}_2^2\, \vecbf{\ell}_3^2\, 
(\vecbf{\ell}_{123} {-} \vecbf{q})^2},
\\[0.5 em]
\vecbf{B}_2  &= e^{3\epsilon\gamma_E}\,
\int{\dd^d\ell_1 \dd^d\ell_2 \dd^d\ell_3 \over \pi^{3d/2}}\,
{(\vecbf{q}^2)^{4 + (1- 3d)/2} \over (\pm\ell_{23}^z)\, \vecbf{\ell}_1^2\, \vecbf{\ell}_2^2\, \vecbf{\ell}_3^2\, 
(\vecbf{\ell}_{123} {-} \vecbf{q})^2}, \label{eq:B2}
\\[0.5 em]
\vecbf{D}_1 &= e^{3\epsilon\gamma_E}\,
\int{\dd^d\ell_1 \dd^d\ell_2 \dd^d\ell_3 \over \pi^{3d/2}}\,
{(\vecbf{q}^2)^{5 + (1- 3d)/2} \over (\pm\ell_3^z)\,\vecbf{\ell}_1^2\, \vecbf{\ell}_2^2\, \vecbf{\ell}_3^2\, (\vecbf{\ell}_{13} {-} \vecbf{q})^2\, 
(\vecbf{\ell}_{23} {-} \vecbf{q})^2}.
\end{align}
We have suppressed the sign superscript since these integrals are independent of the sign of the single linear propagator.
A direct evaluation using the one-loop integrals in \eqref{pm-1-loop} results in
\begin{align}\label{}
\vecbf{B}_1  &= e^{3\epsilon\gamma_E}\,
\frac{i \sqrt{\pi}\, \Gamma(-\epsilon)\, \Gamma(3\epsilon)\, \Gamma(1-3\epsilon)\,\Gamma^3({1/2} - \epsilon)}
{\Gamma (1-4\epsilon)\, \Gamma({3/2} - 3\epsilon)}
\nonumber\\
&= -\frac{i}{3} \pi^{3/2} \bigg[\frac{2}{\epsilon^2} + \frac{12}{\epsilon} + 72 - \frac{5\pi^2}{2}\bigg]
+ \mathcal{O}(\epsilon),
\\[0.5 em]
\vecbf{B}_2  &= e^{3\epsilon\gamma_E}\,
\frac{i \sqrt{\pi}\, \Gamma(3\epsilon)\, \Gamma^2({1/2}-2\epsilon)\, \Gamma^4({1/2}-\epsilon)}{\Gamma(1-4\epsilon)\,\Gamma^2 (1-2\epsilon)}
\nonumber\\
&= \frac{1}{3} i \pi ^{7/2} \bigg[
\frac{1}{\epsilon} + 16\log{2} + \epsilon\left(\frac{7}{4}\,\pi^2 +128 \log^2 {2}\right)
\bigg] + \mathcal{O}(\epsilon^2),
\\[0.5 em]
\vecbf{D}_1 &= e^{3\epsilon\gamma_E}\,
\frac{i \sqrt{\pi}\, \Gamma(-3\epsilon)\, \Gamma (-\epsilon)\, \Gamma(1+ 3\epsilon)\,\Gamma^2(1/2 + \epsilon)\, \Gamma^4(1/2 - \epsilon)}{\Gamma^2(1-2\epsilon)\, \Gamma(-4 \epsilon)\, \Gamma(1 + 2\epsilon)}
\nonumber\\
&= -\frac{4}{3} i \pi ^{7/2} \bigg[\frac{1}{\epsilon} + 4\log{2} + \epsilon\left(\frac{3}{4}\,\pi^2 +8 \log^2{2}\right)\bigg]
+ \mathcal{O}(\epsilon^2).
\end{align}
Next, we find four static master integrals in the presence of two linear propagators \cite{Dlapa:2021npj, Dlapa:2021vgp}:
\begin{align}\label{eq:B3B4}
\vecbf{B}^{\pm}_{3} &= e^{3\epsilon\gamma_E}
\int{\dd^d\ell_1 \dd^d\ell_2 \dd^d\ell_3 \over \pi^{3d/2}}\,
{1 \over (\ell_1^z)\, (\pm\ell_2^z)}\,
{(\vecbf{q}^2)^{5 - 3d/2} \over \vecbf{\ell}_1^2\, \vecbf{\ell}_2^2\, \vecbf{\ell}_3^2\, (\vecbf{\ell}_{123} - \vecbf{q})^2},
\\[0.5 em]
%%%
\vecbf{B}^{\pm}_{4} &= e^{3\epsilon\gamma_E}\,
\int{\dd^d\ell_1 \dd^d\ell_2 \dd^d\ell_3 \over \pi^{3d/2}}\,
{1 \over (\ell_1^z)\, (\pm\ell_{12}^z)}\,
{(\vecbf{q}^2)^{5 - 3d/2} \over \vecbf{\ell}_1^2\, \vecbf{\ell}_2^2\, \vecbf{\ell}_3^2\, (\vecbf{\ell}_{123} - \vecbf{q})^2}.
\end{align}
The following analytic results in terms of hypergeometric functions ${}_pF_q$ have been computed for the results presented in~\cite{Dlapa:2021npj,Dlapa:2021vgp}.
An extended analytic derivation is given in~\cite{Dlapa:2023hsl}, which we have generalized to higher loops in App.~\ref{app:deformation}
\begingroup
\allowdisplaybreaks
\begin{align}
\vecbf{B}^{\pm}_{3}
&= e^{3 \epsilon \gamma_E} \frac{\Gamma_{1 / 2 + 3 \epsilon} \Gamma_{1 / 2 - 3 \epsilon}^2 \Gamma_{1 / 2 - \epsilon}^2}{\Gamma_{1 - 2 \epsilon}} \left[ - \frac{\pi \Gamma_{1 / 2 - 2 \epsilon} \Gamma_{- \epsilon}^2}{\Gamma_{1 / 2 - 4 \epsilon} \Gamma_{1 / 2 - 3 \epsilon}^2} \right.
\nn\\
&\quad \mp \frac{2 \pi}{1 - 4 \epsilon} \frac{\csc (2 \pi \epsilon)}{\Gamma_{1 - 6 \epsilon}} \, _3F_2\big( \tfrac{1}{2} {-} 3\epsilon, \tfrac{1}{2} {-} 3\epsilon, \tfrac{1}{2} {-} 2\epsilon ; 1 {-} 6\epsilon, \tfrac{3}{2} {-} 2\epsilon ; 1\big)
\nn\\
&\quad\left. \mp  \frac{2 \Gamma_{1 / 2 - \epsilon}^2 \Gamma_{- 2 \epsilon}}{\Gamma_{1 - 4 \epsilon} \Gamma_{1 / 2 - 3 \epsilon}^2} \, _4F_3 \big( \tfrac{1}{2}, 1, \tfrac{1}{2} {-} \epsilon, \tfrac{1}{2} {-} \epsilon ; \tfrac{3}{2}, 1 {-} 4\epsilon, 1 {+} 2\epsilon ; 1 \big) \right],
\\[0.5 em]
\vecbf{B}^{\pm}_{4}
&= e^{3 \epsilon \gamma_E} \frac{\Gamma_{1 / 2 + 3 \epsilon} \Gamma_{1 / 2 - 3 \epsilon} \Gamma_{1 / 2 - \epsilon}^2 \Gamma_{- 2 \epsilon}}{\Gamma_{1 - 2 \epsilon}} \left[ - \frac{\pi\Gamma_{1 / 2 - 2 \epsilon} \Gamma_{- \epsilon}^2}{\Gamma_{1 / 2 - 4 \epsilon} \Gamma_{1 / 2 - 3 \epsilon} \Gamma_{- 2 \epsilon}} \right.
\nn\\
&\quad \mp \frac{\pi}{\epsilon} \frac{\sec (\pi \epsilon)}{\Gamma_{1 / 2 - 5 \epsilon}} \, _3F_2\big( \tfrac{1}{2} {-} 3\epsilon, - 2\epsilon, - \epsilon; \tfrac{1}{2} {-} 5\epsilon, 1 {-} \epsilon; 1 \big) \nn\\
&\quad\left. \pm \frac{2^{2 + 6 \epsilon} \pi \Gamma_{- 1 / 2 - \epsilon}}{\Gamma_{1 / 2 - 3 \epsilon} \Gamma_{1 / 2 - 2 \epsilon} \Gamma_{- \epsilon}} \, _4F_3\big(\tfrac{1}{2}, 1, 1 {-} 2 \epsilon, \tfrac{1}{2} {-} \epsilon; \tfrac{3}{2}, 1 {-} 4 \epsilon, \tfrac{3}{2} {+} \epsilon; 1\big) 
\right],
\end{align}
\endgroup
with $\Gamma_{a}$ being a shorthand notation of $\Gamma (a)$.
%A derivation of these formulae is given in App.~\ref{sec:appA}.
%We include there the proof of the relation $\vecbf{B}_3^+=2\vecbf{B}_4^+$.
Performing the Laurent expansions in $\epsilon$ for the first few orders is surprisingly tricky.\footnote{An alternative approach is given by \emph{multi-sum} techniques, see \cite{Ablinger:2010pb,Blumlein:2011kef,Schneider:2013zna,Blumlein:2021pgo}.}
We numerically evaluated the expansion coefficients and conjecture the following analytic expressions using \textit{Mathematica's} built-in implementation of the PSLQ algorithm \texttt{FindIntegerNullVector}:
\begin{align}
%4
\vecbf{B}^{-}_{3} 
&=  -\pi^{5/2}\bigg[
  \begin{multlined}[t]
    \frac{1}{\epsilon^2} - \frac{6\log(2)}{\epsilon} - \frac{1}{12} (17 \pi^2 -216 \log^2(2))\\
    +\frac{1}{2}\left(17\pi^2\log(2)-72\log^3(2)-606\zeta(3)\right)\epsilon\bigg] + \mathcal{O}(\epsilon^2)\,,
  \end{multlined}
\\[0.35 em]
%7
\vecbf{B}^{+}_{3}
&= 2\vecbf{B}^{+}_{4} = -\pi^{5/2}\bigg[
  \begin{multlined}[t]
    \frac{1}{\epsilon^2} - \frac{6\log(2)}{\epsilon} + \frac{1}{12}(7 \pi^2 + 216 \log^2(2))\\
	-\frac{1}{2}\left(7\pi^2\log(2)+72\log^3(2)+158\zeta(3)\right)\epsilon\bigg]+\mathcal{O}(\epsilon^2)\,,
  \end{multlined}
\\[0.35 em]
%2
\label{eqn:B4m}
\vecbf{B}^{-}_{4}
&= - {3 \over 2} \pi^{5/2} \bigg[
  \begin{multlined}[t]
    \frac{1}{\epsilon^2} - \frac{6}{\epsilon}\log(2) - \frac{3}{4}(\pi^2-24\log^2(2))\\
    -\frac{1}{6}\left(-27\pi^2\log(2)+216\log^3(2)+1370\zeta(3)\right)\epsilon\bigg] + \cO(\epsilon^2)\,.
  \end{multlined}
\end{align}
More details about this reconstruction will be given in Sec.~\ref{sec:setup}.
We have also checked that the above results satisfy the relation
\begin{align}\label{eq:B3B4id}
\vecbf{B}^{+}_{3} + \vecbf{B}^{-}_{3} = \vecbf{B}^{+}_{4} + \vecbf{B}^{-}_{4}
&= 
{\vecbf{A}_{011}  \over \pi^{d}}\,
{(2\pi i)^2 \over 2}\,
\int {\dd^{d-1}\ell_1^\perp \dd^{d-1} \ell_2^\perp\,e^{2\epsilon\gamma_E}\,(\vecbf{q}^2)^{4 - 3d/2}
\over 
(\vecbf{\ell}_1^\perp)^2\, (\vecbf{\ell}_2^\perp)^2\,
[(\vecbf{\ell}_{12}^\perp - \vecbf{q})^2]^{(4-d)/2}}
\nonumber\\[0.35 em]
&= 
-\frac{2\pi\,e^{3\gamma_E \epsilon}\, \Gamma({1}/{2}-2\epsilon)\, \Gamma^2({1/2} - \epsilon)\, \Gamma^2(-\epsilon)\, \Gamma({1/2} + 3\epsilon)}{\Gamma({1/2} - 4\epsilon)\, \Gamma(1-2\epsilon)},
\end{align}
which follows from the fact that the combination of linear propagators with different signs forms a maximal cut of all linear propagators.

Finally, let us consider static integrals with three linear propagators:
\begin{align}\label{}
\vecbf{B}_5^{\pm\pm}  &= e^{3\epsilon\gamma_E}\,
\int {\dd^{d} \ell_1 \dd^{d} \ell_2 \dd^{d} \ell_3 \over \pi^{3d/2}}\,
{(\vecbf{q}^2)^{4 + (3- 3d)/2}
\over (\ell_1^z) (\pm\ell_{12}^z) (\mp\ell_3^z)\, \vecbf{\ell}_1^2\, \vecbf{\ell}_2^2\, \vecbf{\ell}_3^2\, 
(\vecbf{\ell}_1 {+} \vecbf{\ell}_2 {+} \vecbf{\ell}_3 {-} \vecbf{q})^2},
\\[0.5 em]
\vecbf{B}_6^{\pm\pm}  &= e^{3\epsilon\gamma_E}\,
\int {\dd^{d} \ell_1 \dd^{d} \ell_2 \dd^{d} \ell_3 \over \pi^{3d/2}}\,
{(\vecbf{q}^2)^{4 + (3- 3d)/2} 
\over (\ell_1^z) (\pm\ell_2^z) (\pm\ell_3^z)\, \vecbf{\ell}_1^2\, \vecbf{\ell}_2^2\, \vecbf{\ell}_3^2\, 
(\vecbf{\ell}_1 {+} \vecbf{\ell}_2 {+} \vecbf{\ell}_3 {-} \vecbf{q})^2}.
\end{align}
%Like the two-loop integrals in \eqref{eq:K2_6} and \eqref{eq:K2_7}, these integrals can be analytically evaluated using a symmetrization trick to express them in terms of 2-dimensional \emph{banana} integrals (see App.~\ref{sec:appA}):
We find that they fulfil the following non-trivial relations:
\begin{align}\label{}
\begin{aligned}
&
\vecbf{B}_5^{+-}  = 3\,\vecbf{B}_5^{++}, \quad
\vecbf{B}_5^{-+}  = 5\,\vecbf{B}_5^{++} , \quad
\vecbf{B}_5^{--}  = 3\,\vecbf{B}_5^{++},
\\[0.3 em]
&
\vecbf{B}_6^{++}  = 6\,\vecbf{B}_5^{++}, \quad
\vecbf{B}_6^{+-}  = 2\,\vecbf{B}_5^{++}, \quad
\vecbf{B}_6^{-+}  = 2\,\vecbf{B}_5^{++}, \quad
\vecbf{B}_6^{--}  = 2\,\vecbf{B}_5^{++},
\end{aligned}
\end{align}
and
\begin{align}\label{eq:B56ids}
\vecbf{B}_5^{++}  &= {(2\pi i)^3 \over 24}\, {(\vecbf{q}^2)^{4 - 3(d-1)/2} \over \pi^{3d/2}}\,
\int  {\dd^{d-1}\ell_1^\perp \dd^{d-1}\ell_2^\perp \dd^{d-1}\ell_3^\perp\, e^{3\epsilon\gamma_E}    
\over (\vecbf{\ell}_1^\perp)^2\, (\vecbf{\ell}_2^\perp)^2\, (\vecbf{\ell}_3^\perp)^2\, (\vecbf{\ell}_{123}^\perp - \vecbf{q})^2}
\nonumber\\
&= 
- {i \pi ^{3/2} e^{3\gamma_E  \epsilon}\, \Gamma^4(-\epsilon)\, \Gamma(3 \epsilon +1)  \over  3 \Gamma(-4 \epsilon )}
\nonumber\\
&=
{4i \over 3}\,\pi^{3/2}\,
\bigg[
\frac{1}{\epsilon ^3}
-\frac{\pi ^2}{4 \epsilon }
-29 \zeta (3)
\bigg]
+ \mathcal{O}(\epsilon).
\end{align}
They satisfy
\begin{align}\label{}
\vecbf{B}_j^{++} + \vecbf{B}_j^{+-} + \vecbf{B}_j^{-+} + \vecbf{B}_j^{--}  &= 
{(2\pi i)^3 \over 2}\, {(\vecbf{q}^2)^{(11 - 3d)/2} \over \pi^{3d/2}}\,
\int  {\dd^{d-1}\ell_1^\perp \dd^{d-1}\ell_2^\perp \dd^{d-1}\ell_3^\perp\, e^{3\epsilon\gamma_E}    
\over (\vecbf{\ell}_1^\perp)^2\, (\vecbf{\ell}_2^\perp)^2\, (\vecbf{\ell}_3^\perp)^2\, (\vecbf{\ell}_{123}^\perp - \vecbf{q})^2},
\end{align}
for $j=5,6$ respectively.
These relations are following from the fact that the combination of linear propagators with different signs yields a maximal cut of all linear propagators.
This completes the set of all static master integrals for the conservative, non-spinning contributions at $\mathcal{O}(G^4)$.
The results for $\vecbf{B}_1$, $\vecbf{B}_2$, $\vecbf{D}_1$, $\vecbf{B}_5$, and $\vecbf{B}_6$ are to our knowledge presented for the first time here.

%**************************************************
\subsection{5PM: Four loops}\label{sec:pm-int-4L}

We pick a representative set of integrals which are likely to appear as static master integrals in up-coming computations for the conservative dynamics at 5PM order.
Here we choose to study the most typical one, the four-loop \emph{banana} topology, as a representative to test our numerical methods:
\begin{align*}\label{}
\begin{aligned}
\begin{tikzpicture}[scale=0.7] 
  \draw[line width=0.8pt] (0,0) arc (10:170:2 and 1.2);
  \draw[line width=0.8pt] (0,0) arc (10:170:2 and 0.6);
  \draw[line width=0.8pt] (-4.0,0) -- (0.8,0);
  \draw[line width=0.8pt] (0,0) arc (-10:-170:2 and 0.6);
  \draw[line width=0.8pt] (0,0) arc (-10:-170:2 and 1.2);
  \draw[xshift=0cm, -, line width=1.2pt] (0,0) -- (0.8,0);
  \draw[xshift=-39.2mm, -, line width=1.2pt] (0,0) -- (-0.8,0); 
\end{tikzpicture}\,.
\end{aligned}
\end{align*}
Let us first consider the simplest case without any linear propagator:
\begin{align}\label{}
\vecbf{M}_0 &= e^{4\epsilon\gamma_E}\,
\int  {\dd^{d}\ell_1 \dd^{d}\ell_2 \dd^{d}\ell_3 \dd^{d}\ell_4  \over  \pi^{2d}}\,
{(\vecbf{q}^2)^{5-2d} 
\over  \vecbf{\ell}_1^2\, \vecbf{\ell}_2^2\, \vecbf{\ell}_3^2\, \vecbf{\ell}_4^2\, 
(\vecbf{\ell}_1 {+} \vecbf{\ell}_2 {+} \vecbf{\ell}_3 {+} \vecbf{\ell}_4 {-} \vecbf{q})^2}.
\end{align}
Its analytic result, obtained once more via iterative application of the one-loop bubble formula~\eqref{pm-1-loop}, is given by
\begin{align}\label{}
\vecbf{M}_0
&= 
e^{4 \gamma_E \epsilon}\, \frac{\Gamma^5({1/2}-\epsilon)\, \Gamma(4\epsilon -1)}{\Gamma({5/2}-5\epsilon)}
\nonumber\\
&= 
-\frac{\pi ^2}{3 \epsilon }
-\frac{52 \pi ^2}{9}
+ \frac{1}{27} \pi ^2 \left(33 \pi ^2-1924\right) \epsilon
+ \mathcal{O}(\epsilon^2).
\end{align}
We consider a generalization with a single linear propagator:
\begin{align}\label{}
\vecbf{M}_1 = 
e^{4\epsilon\gamma_E} \int  {\dd^{d}\ell_1 \dd^{d}\ell_2 \dd^{d}\ell_3 \dd^{d}\ell_4  \over \pi^{2d}}
{(\vecbf{q}^2)^{11/2-2d}
\over  (\pm\ell_1^z)\, \vecbf{\ell}_1^2\, \vecbf{\ell}_2^2\, \vecbf{\ell}_3^2\, \vecbf{\ell}_4^2\, 
(\vecbf{\ell}_{1234} - \vecbf{q})^2}.
\end{align}
Similarly, its analytic form can be obtained using the one-loop bubble integral:
\begin{align}\label{}
\vecbf{M}_1 &= e^{4\gamma_E\epsilon}\, 
{i \sqrt{\pi }\, \Gamma({3/2} - 4\epsilon)\, \Gamma^4({1/2} - \epsilon)\, \Gamma(-\epsilon)\, \Gamma(4\epsilon - 1/2)
\over  \Gamma({3/2}-5 \epsilon)\, \Gamma(2-4 \epsilon)}
\\
&= {i\pi^3}
\bigg(
\frac{2}{\epsilon }  
+ 4 (7-\log 2)
+ \big(3(104 + \pi^2) + 4 (\log2 - 14) \log2\big) \epsilon 
\bigg)
+{\cal O} (\epsilon^2).
\end{align}
Adding another linear propagator, we consider the following integrals
\begin{align}\label{}
\vecbf{M}_2^\pm  &= \int 
{\dd^{d}\ell_1 \dd^{d}\ell_2 \dd^{d}\ell_3 \dd^{d}\ell_4   \over \pi^{2d}}
{(\vecbf{q}^2)^{7-2d} \,e^{4\epsilon\gamma_E} \over (\ell_1^z) (\pm\ell_2^z)\,
\vecbf{\ell}_1^2\, \vecbf{\ell}_2^2\, \vecbf{\ell}_3^2\, \vecbf{\ell}_4^2\, 
(\vecbf{\ell}_{1234} {-} \vecbf{q})^2}\,,
\\[0.35 em]
\vecbf{M}_3^\pm  &= \int 
{\dd^{d}\ell_1 \dd^{d}\ell_2 \dd^{d}\ell_3 d^{d}\ell_4   \over \pi^{2d}}
{(\vecbf{q}^2)^{7-2d} \,e^{4\epsilon\gamma_E} \over (\ell_1^z) (\pm\ell_{12}^z)\,
\vecbf{\ell}_1^2\, \vecbf{\ell}_2^2\, \vecbf{\ell}_3^2\, \vecbf{\ell}_4^2\, 
(\vecbf{\ell}_{1234} {-} \vecbf{q})^2}\,,
\end{align}
which have the analytic solution
\begingroup
\allowdisplaybreaks
\begin{align}
\vecbf{M}_2^\pm
&= e^{4 \epsilon \gamma_E} \frac{\Gamma_{1 - 4 \epsilon}^2 \Gamma_{1 / 2 - \epsilon}^3 \Gamma_{4 \epsilon}}{\Gamma_{3 /  2 - 3 \epsilon}} \left[ - \frac{\pi\, \Gamma_{1 - 3 \epsilon} \Gamma_{- \epsilon}^2}{\Gamma_{1 - 5 \epsilon} \Gamma_{1 - 4 \epsilon}^2} \right.
\nn\\
&\quad \pm \frac{\pi}{1 - 3 \epsilon} \frac{\sec (3 \pi \epsilon)}{\Gamma_{2 - 8 \epsilon}}~{}_3F_2 \left( 1 - 4 \epsilon, 1 - 4 \epsilon, 1 - 3 \epsilon;  2 - 8 \epsilon, 2 - 3 \epsilon ; 1 \right)
\nn\\
&\quad\left. \mp \frac{2 \Gamma_{1 / 2 - 3 \epsilon} \Gamma_{1 / 2 - \epsilon}^2}{\Gamma_{3 / 2 - 5 \epsilon} \Gamma_{1 - 4 \epsilon}^2}~{}_4F_3 \big( \tfrac{1}{2}, 1, \tfrac{1}{2} - \epsilon, \tfrac{1}{2} - \epsilon; \tfrac{3}{2}, \tfrac{3}{2} - 5 \epsilon, \tfrac{1}{2} + 3 \epsilon; 1 \big) \right]\,,
\\[0.5 em]
%%%
\vecbf{M}_3^\pm
&=
e^{4 \epsilon \gamma_E} \frac{\Gamma_{1 - 4 \epsilon} \Gamma_{1 / 2 - \epsilon}^3 \Gamma_{- 2 \epsilon} \Gamma_{4 \epsilon}}{\Gamma_{3 / 2 - 3 \epsilon}} \left[ - \frac{\pi\, \Gamma_{1 - 3 \epsilon} \Gamma_{- \epsilon}^2}{\Gamma_{1 - 5 \epsilon} \Gamma_{1 - 4 \epsilon} \Gamma_{- 2 \epsilon}} \right.
\nn\\
&\quad
\mp \frac{\pi}{\epsilon} \frac{\sec (\pi \epsilon)}{\Gamma_{1 - 6 \epsilon}}~{}_3F_2 \big( 1 - 4 \epsilon, - 2 \epsilon, - \epsilon; 1 - 6 \epsilon, 1 - \epsilon; 1 \big)
\nn\\
&\quad\left.
\pm \frac{2 \Gamma_{3 / 2 - 3 \epsilon} \Gamma_{- 1 / 2 - \epsilon} \Gamma_{1 / 2 - \epsilon}}{\Gamma_{3 / 2 - 5 \epsilon} \Gamma_{1 - 4 \epsilon} \Gamma_{-2 \epsilon}}~{}_4F_3 \big( \tfrac{1}{2}, 1, \tfrac{3}{2} - 3 \epsilon, \tfrac{1}{2} - \epsilon; \tfrac{3}{2}, \tfrac{3}{2} - 5 \epsilon, \tfrac{3}{2} + \epsilon; 1 \big)
\right]\,.
\end{align}
\endgroup
%We derive these results in App.~\ref{sec:appA}.
These results can be obtained in a similar way as the 3-loop integrals $\vecbf{B}_3$ and $\vecbf{B}_4$ in Eq.~\eqref{eq:B3B4}.
Since the hypergeometric functions start contributing only at the third order in $\epsilon$ we can analytically perform the series expansion up to that order.
We realized that by multiplying this series by $(1-6\epsilon)$ leads to a uniform transcendental result.
This allowed us then to conjecture the coefficient at $\cO(\epsilon^0)$ via an integer relation algorithm (see Sec.~\ref{sec:setup} for more details):
\begin{align}
  (1-6\epsilon)\vecbf{M}_2^+ &= -\frac{\pi^2}{2}\left(\frac{1}{\epsilon^3}-\frac{\pi^2}{\epsilon}-\frac{256\zeta(3)}{3}\right) + \cO(\epsilon)\,,\\
  (1-6\epsilon)\vecbf{M}_2^- &= -\frac{\pi^2}{2}\left(\frac{1}{\epsilon^3}-\frac{5\pi^2}{3\epsilon}-\frac{400\zeta(3)}{3}\right) + \cO(\epsilon)\,,\\
  (1-6\epsilon)\vecbf{M}_3^- &= -\frac{\pi^2}{4}\left(\frac{3}{\epsilon^3}-\frac{13\pi^2}{3\epsilon}-352\zeta(3)\right) + \cO(\epsilon)\,,
\end{align}
and $\vecbf{M}_3^+ = {1 \over 2} \vecbf{M}_2^+$.
%The latter relation can be proved analogously to the two-loop case.
We have further checked that the above results satisfy the relations
\begin{align}
\vecbf{M}_j^{+} + \vecbf{M}_j^{-} 
&= 
{\Gamma^3({1/2} - \epsilon)\, \Gamma(2\epsilon) \over \Gamma({3/2} - 3\epsilon)}
{e^{4\epsilon\gamma_E} \over \pi^{d}}\,
{(2\pi i)^2 \over 2}\,
\int {\dd^{d-1}\ell_1^\perp \dd^{d-1} \ell_2^\perp\,(\vecbf{q}^2)^{5 - 2d}
\over 
(\vecbf{\ell}_1^\perp)^2\, (\vecbf{\ell}_2^\perp)^2\,
[(\vecbf{\ell}_{12}^\perp - \vecbf{q})^2]^{3-d}}
\nonumber\\[0.35 em]
&=
- e^{4\gamma_E \epsilon}\, {2\pi\,  \Gamma(1-3\epsilon)\, \Gamma^3({1/2} - \epsilon)\, \Gamma^2(-\epsilon)\, \Gamma(4\epsilon)  \over  \Gamma(1-5\epsilon)\, \Gamma({3/2} - 3\epsilon)},
\end{align}
with $j=2,3$.

We will not consider any integral with three linear propagators.
Finally, we consider an integral with four linear propagators
\begin{align}\label{}
\vecbf{M}_4^{\pm\pm\pm}  &= \int 
{\dd^{d}\ell_1 \dd^{d}\ell_2 \dd^{d}\ell_3 \dd^{d}\ell_4   \over \pi^{2d}}
{(\vecbf{q}^2)^{7-2d} \,e^{4\epsilon\gamma_E} \over (\ell_1^z) (\pm\ell_{12}^z) (\pm\ell_{123}^z) (\mp\ell_{4}^z)\,
\vecbf{\ell}_1^2\, \vecbf{\ell}_2^2\, \vecbf{\ell}_3^2\, \vecbf{\ell}_4^2\, 
(\vecbf{\ell}_{1234} {-} \vecbf{q})^2}\,.
\end{align}
%Like the two and three loop cases, these integrals can be evaluated in terms of 2d banana integrals (see App.~\ref{sec:appA} for the derivations)
They fulfill the following relations:
\begin{align}\label{}
\begin{aligned}
\vecbf{M}_4^{++-}  &= 4\vecbf{M}_4^{+++}\,,
~~
\vecbf{M}_4^{+-+}  = 9\vecbf{M}_4^{+++}\,,
~~
\vecbf{M}_4^{+--}  = 6\vecbf{M}_4^{+++}\,,
~~
\vecbf{M}_4^{-++}  = 9\vecbf{M}_4^{+++}\,,
\\[0.3 em]
\vecbf{M}_4^{-+-}  &=  16\vecbf{M}_4^{+++}\,,
~~
\vecbf{M}_4^{--+}  = 11\vecbf{M}_4^{+++}\,,
~~
\vecbf{M}_4^{---}  = 4\vecbf{M}_4^{+++}\,.
\end{aligned}
\end{align}
and
\begin{align}\label{eq:M4ids}
\vecbf{M}_4^{+++}  &= {(\vecbf{q}^2)^{7-2d} \over \pi^{2d}}\, {(2\pi i)^4 \over 120} \int
{\dd^{d-1}\ell_1 \dd^{d-1}\ell_2 \dd^{d-1}\ell_3 \dd^{d-1}\ell_4 \,e^{4\epsilon\gamma_E}  \over
(\vecbf{\ell}_1^\perp)^2\, (\vecbf{\ell}_2^\perp)^2\, (\vecbf{\ell}_3^\perp)^2\, (\vecbf{\ell}_4^\perp)^2\,
(\vecbf{\ell}_{1234}^\perp - \vecbf{q})^2}
\\[0.35 em]
&=  e^{4\epsilon\gamma_E}\, {2 \pi^2\, \Gamma^5(-\epsilon)\, \Gamma (4 \epsilon +1) \over 15\Gamma (-5\epsilon )}
\\[0.3 em]
&=
\frac{2 \pi ^2}{3}
\bigg(
\frac{1}{\epsilon^4}
-\frac{\pi^2}{3 \epsilon^2}
-\frac{184 \zeta (3)}{3 \epsilon}
-\frac{43\pi^4}{45}
\bigg)
+ \mathcal{O}(\epsilon)\,.
\end{align}
They furthermore satisfy the following non-trivial relation
\begin{align}\label{}
\vecbf{M}_4^{+++}  + \vecbf{M}_4^{++-} &+ \vecbf{M}_4^{+-+}
+ \vecbf{M}_4^{+--} + \vecbf{M}_4^{-++} + \vecbf{M}_4^{-+-} + \vecbf{M}_4^{--+} + \vecbf{M}_4^{---} 
\nonumber\\
&= e^{4\epsilon\gamma_E}\, {(\vecbf{q}^2)^{7-2d} \over \pi^{2d}}\, {(2\pi i)^4 \over 2}
\int {\dd^{d-1}\ell_1^\perp \dd^{d-1}\ell_2^\perp \dd^{d-1}\ell_3^\perp d^{d-1}\ell_4^\perp   \over
(\vecbf{\ell}_1^\perp)^2\, (\vecbf{\ell}_2^\perp)^2\, (\vecbf{\ell}_3^\perp)^2\, (\vecbf{\ell}_4^\perp)^2\,
(\vecbf{\ell}_{1234}^\perp {-} \vecbf{q})^2}\,.
\end{align}
The analytic results for the integrals $\vecbf{M}_2$, $\vecbf{M}_3$, and $\vecbf{M}_4$ are to our knowledge for the first time presented here.
A derivation based on a symmetrization trick can be found in App.~\ref{sec:appA}.
All of these integrals are vital for the conservative contributions to the binary dynamics at $\cO(G^5)$.

%%%%%%%%%%%%%%%%%%%%%%%%%%%%%%%%%%%%%%%%%%%%%%%%%%
\section{Numerical methods and results}\label{sec:numerical}
%%%%%%%%%%%%%%%%%%%%%%%%%%%%%%%%%%%%%%%%%%%%%%%%%%

In this section we present a framework to numerically evaluate dimensionally-regularized multi-loop integrals, with a special focus on the integrals introduced in Sec.~{\ref{sec:theory}}.
We start by discussing some background material, in which we explain the two main steps in our computation: sector decomposition and Monte Carlo integration. 
We then present a neural network method to sample the phase space. 
Next, we detail our explicit software setup and an analysis of the desired precision for numerical methods having in mind analytical reconstruction methods.
We finish by presenting a comparison of NNs and \vegas numerical integral solvers applied to PM boundary integrals.

\subsection{Prerequisites}
The numerical evaluation consists of two steps:
First, the integral is decomposed into different sectors in order to write it as a Laurent series in the dimensional regularization parameter~$\epsilon$, where each series coefficient is expressed as a purely numerical integral.
Methods that implement such an expansion are called \emph{sector decomposition}~\cite{Prokhorenko:2007yy,Roth:1996pd,Binoth:2000ps,Heinrich:2008si}.
Second, we numerically evaluate these integrals using Monte Carlo-based methods.
We start by presenting two different codes for sector decomposition, \fiesta~\cite{Smirnov:2008py,Smirnov:2009pb,Smirnov:2013eza,Smirnov:2015mct,Smirnov:2021rhf} and \pysecdec~\cite{Carter:2010hi,Borowka:2012yc,Borowka:2015mxa,Borowka:2017idc},
followed by a review of two different Monte Carlo methods for the evaluation of the integrals: the widely-used \vegas algorithm and a novel method based on neural networks.

%**************************************************
\subsubsection{Sector decomposition: \fiesta and \pysecdec}\label{sec:fiesta-pysecdec}

Sector decomposition techniques date back to the proof of the BPHZ theorem~\cite{Hepp} and have been used to isolate both infrared (IR) and ultraviolet (UV) singularities of Feynman integrals~\cite{Roth:1996pd}.
Different strategies for sector decompositions~\cite{Binoth:2000ps,Bogner:2007cr,Smirnov:2008py} can lead to different number of sectors and distinct structure of the poles in the regulator $\epsilon$.
Also, the convergence properties of the algorithm can vary significantly among different strategies~\cite{Bogner:2007cr, Kaneko:2009qx}.
The general idea of sector decomposition (see e.g. \cite{Heinrich:2008si} for a review) is to split the integration region iteratively into smaller pieces, such that overlapping singularities (a denominator evaluates to zero for a set of integration variables $x_i\rightarrow 0$) are factorized.
This is always possible, and proven to terminate for appropriate strategies due to homogeneousness properties of (Feynman) parametrized integrals.
Having arrived at such a factorized form the extraction of poles is trivialized and a Laurent series in $\epsilon$ can be extracted to any desired order.

Further singularities of the integrand at other points (surfaces) away from zero need to be handled by a \emph{contour deformation}~\cite{Soper:1999xk}.
It utilizes a complex deformation dictated by the $i0$ prescription of the (Feynman) propagators.

We have used two different programs to study and automatize the sector decomposition and contour deformation for the PM boundary integrals.
\fiesta was first developed in \cite{Smirnov:2008py} and improved in \cite{Smirnov:2009pb, Smirnov:2013eza}.
Its core algorithms are implemented in C and a Mathematica interface is provided.
\fiesta provides many different strategies for sector decomposition based on work in \cite{Smirnov:2008py, Smirnov:2013eza, Kaneko:2009qx}.

\texttt{SecDec} was developed in both C++ and Fortran \cite{Carter:2010hi, Borowka:2012yc, Borowka:2015mxa} and has a Python interface (\texttt{pySecDec}) \cite{Borowka:2017idc}.
It allows for three different decomposition strategies: {\it iterative}~\cite{Binoth:2000ps, Heinrich:2008si} and two {\it geometric} decomposition methods described in~\cite{Kaneko:2009qx,Borowka:2015mxa} that make use of the \texttt{normaliz}  package \cite{bruns2015power}.
In our tests, we found that the {\it geometric} method described in \cite{Borowka:2015mxa} that makes use of the Cheng-Wu theorem \cite{cheng1987expanding} leads to fewer sectors. It produces for all integrals discussed here the most compact integrand, i.e. allowing for the fastest numerical evaluation at a random phase-space point.
This observation agrees with the analysis made in \cite{Borowka:2015mxa}.
We use this method by default for the rest of this work.

%**************************************************
\subsubsection{Monte Carlo integrators: \vegas family and Neural Networks}\label{sec:NN}

Monte Carlo algorithms estimate an integral $I$ of a function $f(x)$
\begin{equation}
I = \int_\Omega \dd x\, f(x) \,,
\end{equation}
over the domain $\Omega$ by sampling the integrand over $N$ uniformly distributed points $x_i$
\begin{eqnarray}
    I \approx I_{\rm MC} = \frac{V}{N}\sum^N_{i=1} f(x_i) \equiv V \langle f \rangle_x \,,
\end{eqnarray}
where $V$ is the volume of $\Omega$. The brackets represent the average taken with respect to a uniform sampling in the variable $x$. 

{\it Importance sampling} means performing a variable change such that the regions in the phase space with large $|f|$ gain more weight than other regions with small $|f|$.
This decreases the variance~$\sigma$, a measure that we use to estimate the accuracy of the result.
The basic idea is to use a probability density function (PDF) that resembles $f$, $g(x, \theta)\sim f(x)/I$.
It may depend on a {\it nuisance parameter}  $\theta$.
Nuisance parameters are used in the statistics literature to enlarge the parameter space of a theory in order to take into account known unknowns \cite{Dorigo:2020ldg}.
Letting $G(x, \theta)$ be the cumulative distribution of $g$
\begin{equation}
  \dd G(x, \theta) = g(x, \theta) \dd x
\end{equation}
we have
\begin{equation}
    I = \int_\Omega \dd x\, f(x) = \int_{\tilde{\Omega}} \dd G(x, \theta) \, \frac{f(x)}{g(x, \theta)} \simeq V \Biggl\langle \frac{f(x)}{g(x, \theta)} \Biggl\rangle_G\,.
\end{equation}
Putting that in another way, $g$ is the inverse Jacobian determinant $J = |\dd x/\dd G|$. 

The variance of the MC integral is estimated by
\begin{equation} \label{eq:error_scaling}
    \sigma_{\rm MC}^2 = \frac{1}{N-1} \left[ \frac{1}{N}\sum_i \left( \frac{f(x(G_i))}{g(G_i)}\right)^2 - \left(\frac{1}{N}\sum_i \frac{f(x(G_i))}{g(G_i)} \right)^2\right]\,,
\end{equation}
which helps us to understand the effect of importance sampling: $g$ reduces the overall MC variance as good as it resembles $f$, i.e. for an optimal choice of $g(x) = f(x)/I$, in which one already knows $I$, the variance vanishes.
For non-optimal choices it is understood that the better the shape of $g$ resembles $f$ the more the peak regions get suppressed by the Jacobian $J$, reducing the variance of the integrand.
The efficiency of importance sampling is attached to three distinct factors: $g$'s shape should resemble $f$, be invertible, and fast to evaluate (comparable to the cost of evaluating $f$).

In other words, $G(x,\theta)$ is a coordinate transformation.
Sampling uniformly over $G$-coordinates and mapping them to $x$-space (requiring the inverse Jacobian) allows one to reduce the variance. \vegas and \iflow, introduced below, differ by how they construct an importance sampling function $g$.

\paragraph{\vegas:}

\vegas \cite{Lepage:1977sw} is an iterative Monte Carlo scheme that approximates the function $f$ by a histogram function $g$ on a grid.
When computing $d$-dimensional integrals, approximating each dimension by $N$ histogram steps leads to $N^d$ integrand bins. In order to avoid exponential scaling, \vegas assumes  integrand dimensions to be independent, i.e. assumes that
\begin{equation} \label{eq:independence}
    g(x_1,\dots, x_n) = g(x_1)\dots g(x_n)\,,
\end{equation}
leading to $Nd$ integrand bins and therefore a linear scaling.
\vegas is constructed to iteratively refine the binning used to generate the histogram.
After each evaluation of the integrand this refinement is done through a weighting proportional to $J^2f^2$, where $J$ is the Jacobian determinant of the coordinate transformation, evaluated at the previous iteration step.
Hence, the bins get smaller around regions where $|f|$ is larger.

Note that the effectiveness of \vegas depends on the lack of correlation of the integrand among the integration variables, i.e. the assumption underlying Eq.~\eqref{eq:independence}.
For integrands that cannot be factorized, \vegas presents a poor sampling of points \cite{Bendavid:2017zhk}.
Recent versions of \vegas use {\it adaptive stratified sampling} (see \cite{Lepage:2020tgj}) to partially overcome this issue.
Other algorithms, such as FOAM \cite{Jadach_2003} have been proposed for cases in which the integrand is not independent in its components.
FOAM uses an adaptive method to divide the overall phase space into hypercubes taking into account correlations.
Though relatively efficient when dealing with low-dimensional integrals, FOAM becomes inefficient for higher dimensions \cite{Gao:2020vdv}.
Moreover, histogram-based methods lack precision around the edges of the histograms, leading to the so-called {\it edge effects}.
This effect is bypassed by neural networks, which approximate the phase space via splining, as we will discuss now.

\paragraph{\iflow:}
When the independence of components, Eq.\,\eqref{eq:independence}, fails, \vegas generically provides a poor sampling of the phase space and can be inefficient to probe non-diagonal contributions \cite{Bendavid:2017zhk, Gao:2020vdv}, i.e. correlations between different axes in the phase space.
As previously mentioned, a central piece for importance sampling is to find a coordinate map $G$ that satisfy three conditions: its Jacobian resembles the distribution of the integrand $f$, it is invertible and fast to calculate.
Neural networks are able to \emph{learn} an approximation of a given function $f$ that, different from VEGAS, is independent of the axis alignment.
Hence, it captures non-diagonal features.

The basic idea of neural networks is to model (approximate) a function by a concatenation of a number of \emph{layers}.
Each layer depends on the output of the previous layer and some internal parameters.
The NN is then trained on a set of points by tuning these internal parameters.
A trained NN can for example be used as a fast approximation to the original function, or it can provide a (fast) inversion of the original function.

\begin{figure}[ht]
\centering  
  \includegraphics[width=0.9\textwidth]{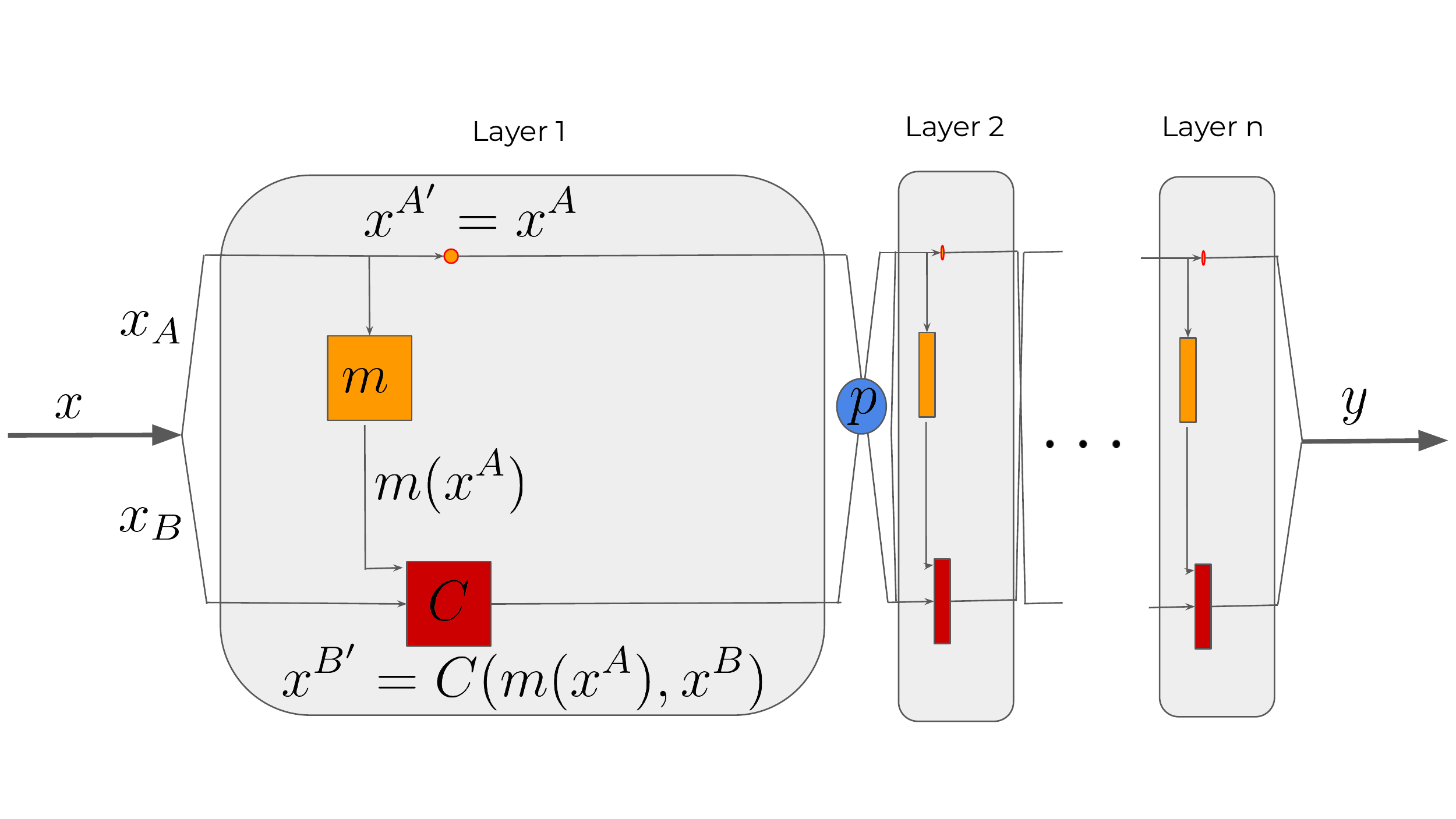}
\caption{\label{fig:flow_scheme}
  \small Normalizing flow scheme.
  We have $n$ coupling layers with coupling transforms $C$ and neural-network functions $m$.
  $x^A$ and $x^B$ are partitions of $x$.
  The $x^A$ goes through a NN transformation $m$ and serve as input together with $x^B$ for a coupling transform $C$.
  The output of $C$, together with $x^A$ serves as input for the second layer, now under a distinct permutation ({\it masking}) $p$.
  See also \cite{muller2019neural}.
}
\end{figure}

Recently, NN architectures that are analytically (i.e. efficiently) invertible were proposed, built through the so-called {\it normalizing flows} technique~\cite{dinh2015nice} (see also~\cite{muller2019neural}).
Normalizing flows make use of {\it coupled layers}, each of which contains itself an efficiently invertible NN. 
The Jacobian matrix of the full transformation is designed to be in an upper-diagonal form whose determinant does not involve the inner neural network function~$m$ (i.e.\,the function that is getting tuned), which only appears in the off-diagonal part.  
Each layer receives a data point~$\vec{x}$ as input from the previous layer.
This point is split into two non-empty subsets $\vec{x}^A$ and $\vec{x}^B$. 
Each layer then outputs a new data point given by $\vec{x}^{A'}=\vec{x}^A$ and $\vec{x}^{B'}=\vec{C}(\vec{m}(x^A),\vec{x}^B)$, where $\vec{C}$ is a coupling function.
For an illustration, see Fig.~\ref{fig:flow_scheme}. 
The coupling function needs to be easily invertible since it appears in the diagonal blocks of the Jacobian:
\begin{equation}
  J=\left| \begin{pmatrix} \vec{1} & 0 \\ \frac{\partial \vec{C}}{\partial \vec{m}}\frac{\partial \vec{m}}{\partial \vec{x}_A} & \frac{\partial \vec{C}}{\partial \vec{x}_B} \end{pmatrix}\right| = \left| \frac{\partial \vec{C}}{\partial \vec{x}_B} \right|\,.
\end{equation}
In Appendix~A of \cite{Gao:2020vdv} some choices for this coupling function are discussed.

The integration algorithm then operates as follows on a batch-by-batch basis:
\begin{enumerate}
    \item Sample uniformly in $G$-space and use the inverted NN to get a point sample in $x$-space;
    \item Make a Monte Carlo estimation for the integral $I$ using both $f(x)$ and $g(x)$;
    \item Update the NN by minimizing a cost function $L(I[g(x)], I[f(x)])$ (that must be provided to \iflow);
    \item Back to item 1 by sampling with the new NN (i.e. updated $G$ and $g$).
\end{enumerate}

NNs have been also used in other ways for the evaluation of (Feynman) integrals, for example to optimize the contour deformation \cite{Winterhalder:2021ngy}.}
Machine-learning-based algorithms have also been shown to overtake \vegas and FOAM for trivial non-factorizable integrands \cite{Bendavid:2017zhk, Klimek:2018mza}. For applications of NNs in Monte Carlo event generation see \cite{Bishara:2019iwh, Gao:2020zvv, Danziger:2021eeg}.

%**************************************************
\subsection{Setup}\label{sec:setup}

For our comparison of the traditional Monte Carlo approach of \vegas and the novel NN implementation of \iflow we used \pysecdec 1.5.2 to construct sector decomposed integrands.
As discussed above we used the \texttt{geometric} decomposition method \cite{Borowka:2015mxa} which produces the most efficient integrands for our purpose.
We optimized the Feynman parametrization by analytically continuing the external data in order to have a positive Symanzik polynomial $\cF$ in cases it was possible.
The analytic continuation that worked for many cases is the following.
Let $\vecbf{u}=(0,0,1)$ such that the linear propagators can be written as $\pm\vecbf{u}\cdot \ell_i-i0$.
The overall power of the $\vecbf{u}$ dependence can be inferred by power counting.
By computing the integral for $\vecbf{u}^2=-1$ instead of $\vecbf{u}^2=1$ many parametrizations have positive Symanzik polynomials and a complex contour deformation is not necessary.
Some technical details related to this are given in App.~\ref{sec:appB}.
The presence of contour deformation typically leads to a poor convergence of the integral.

Some integrals we considered ($\bK^{(++)}_{11;00111}$ and $\vecbf{B}_4^-$) had a technical difficulty related to poles appearing on the boundary of the Symanzik polynomial~$\cF$ when one of the Feynman parameters $x_i\rightarrow 1$.
These are not captured by the standard sector decomposition.
Such poles lead to a poor convergence and in some cases even to erroneous results.
This issue can be resolve by yet another split of the integral into more sectors.
Details about this can be found in \cite{Smirnov:2021rhf} where an option for the newest version of \fiesta was presented that takes care of this issue semi-automatically.
The same paper also discusses the correct treatment for one of our two-loop integrals, $\bK^{(+-)}_{11;00111}$, in detail.
In \pysecdec the same can be achieve by performing the split manually.
In the presence of three or more linear propagators this requires quite some manual work.
We have not observed such issues for integrals where no contour deformation was needed.
For the families $\vecbf{B}_5$/$\vecbf{B}_6$ and $\vecbf{M}_4$ we chose to only numerically integrate one integral per family which has a positive $\cF$ Symanzik polynomial, i.e. $\vecbf{B}_5^{++}$ and $\vecbf{M}_4^{+++}$.
All other integrals from these families can be obtained from the identities in Eqs.~\eqref{eq:B56ids} and \eqref{eq:M4ids}.

For the \vegas integrator we used the default setup of the \pysecdec C++ generator, that makes use of \texttt{CUBA} library \cite{Hahn:2004fe}.
The setup of the \iflow pipeline is more involved. 
\iflow makes use of the TensorFlow library \cite{tensorflow2015-whitepaper}. 
In order to expose \pysecdec's integrand to the TensorFlow interface we have created a TensorFlow operator ($\texttt{op}$\footnote{\href{https://www.tensorflow.org/guide/create_op}{\tt www.tensorflow.org/guide/create\_op}.}) directly from the C++ integrand class.
The \iflow code then takes care of the normalizing flows with the number of (piecewise rational quadratic) coupling layers scaling with the dimension of the integral. 
Each coupling layer has 4 hidden layers, each with 32 nodes and a rectified linear activation function (ReLU).
We also used an Adam optimizer \cite{Kingma:2014vow} and an exponential loss function. 
Each epoch of the NN includes 4096 points sampled. 
We noticed that \iflow results are slightly biased, but introducing a pre-training stage can attenuate this issue.\footnote{Also, different loss functions can lead to different bias according to the likelihood found in the data. We thank Luisa Lucie-Smith for pointing that out to us.}
In practice, this pre-training stage means that we run the NN algorithm until it reaches $50\%$ of the required relative precision and then reset the samplings. 
We have checked that this amount of pre-training reduces the bias in the results to below $2\sigma$, where sigma is the target precision.
A more in-depth study of the source of this bias and a proper way to overcome it is required, though it does not change the overall scaling of the NNs and the conclusions of this work. 
Finally, \iflow stores all sampled points, which may incur into memory issues. For some of the integrals reported in Sec.~\ref{sec:Results}, memory limitation has been an obstacle to \iflow, and similar problems were already reported in \cite{Gao:2020zvv}.\footnote{Since the number of samples used for the actual training is way smaller than the total number of samples, keeping only a representative set for the training is a simple way of overcoming such problems.} 

\paragraph{Precision}
In order to get a feeling for the desired precision, depending on the order in the $\epsilon$ power series expansion, we discuss an example of an \emph{analytic reconstruction} approach based on high-precision results.
For some of the integrals in the previous section we were not able to perform a series expansion to arbitrary order in $\epsilon$ even though we were able to derive the complete analytical result.
The reason is the appearance of hypergeometric functions with arguments depending on $\epsilon$, which are inherently difficult to power expand~\cite{Huber:2005yg,Huber:2007dx}.
Away from the leading order we relied on integer reconstruction algorithms to conjecture an analytic result.
Consider the integral $\vecbf{B}_4^-$ where we presented the series expansion in Eq.\,\eqref{eqn:B4m}.
Assuming uniform transcendental weight we built an ansatz of the form
\begin{equation}
  \frac{\vecbf{B}_4^-}{\pi^{5/2}} = \frac{1}{\epsilon^2} a_1 + \frac{1}{\epsilon}\left( b_1 \pi + b_2 \log(2)\right) + \epsilon^0\left(c_1 \pi^2 + c_2 \pi \log(2) + c_3 \log^2(2)\right) + \cO(\epsilon)\,,
\end{equation}
where we included the set of transcendental numbers $\{\pi,\log(2)\}$ only.
At transcendental weight 3 one also has to include $\zeta(3)$ (and possibly other constants).
The unknown rational coefficients $a_1$, $b_1$, $b_2$, $c_1$, $c_2$, and $c_3$ were then determined via the PSLQ algorithm.
%(using \textit{Mathematica's} \texttt{FindIntegerNullVector} function).
The leading coefficient, $a_1=-3/2$, can in fact be analytically computed since it does not involve derivatives of hypergeometric functions, or can be guessed by eye from a numerical result.
For the coefficients $b_1$ and $b_2$ at $\cO\left(\epsilon^{-1}\right)$ the output of \texttt{FindIntegerNullVector} stabilizes already at a precision of 3 digits.
Finally, for the $c_i$ 11 digits were needed for a stable prediction.
To give confidence in such a conjecture one would like to check it up to a much higher precision.
With the full analytical results at hand, we have checked these results up to a precision of 150 digits.
Looking at this from a different angle sometimes a good guess can work equally well since the correctness of the reconstruction can be justified a posteriori, e.g. in our case by comparing to Post-Newtonian results for physical observables that encapsulate all information about the small velocity limit in which we compute the boundary integrals \cite{Dlapa:2021npj,Dlapa:2021vgp}.

For all series expansions where the full analytic results contains hypergeometric functions we observed similar numbers of digits for a stabilization of the PSLQ algorithm.
Dropping the uniform transcendental constraint and including a bigger set of transcendental numbers would accordingly require higher precision results for a stabilization.
Hence, it can be beneficial to identify uniform transcendental integrals in order to use such a construction.

If such ideas should become useful for results away from the leading and maybe subleading term in $\epsilon$, the precision of numerical integration results needs to exponentially increase.
One improvement in that direction is presented in this paper.
We decided to aim for a relative precision of $\sigma = 10^{-4}$ since it is already sufficient to conjecture analytic results for many of the subleading terms. 
In order to get an idea of the scaling behaviour we also give numbers for a relative precision of $10^{-3}$.

%%%%%%%%%%%%%%%%%%%%%%%%%%%%%%%%%%%%%%%%%%%%%%%%%%
\subsection{Numerical Results}\label{sec:Results}
%%%%%%%%%%%%%%%%%%%%%%%%%%%%%%%%%%%%%%%%%%%%%%%%%%

\newcommand{\nprint}[1]{\numprint{#1}}
\newcommand{\nprintbf}[1]{\bf{\numprint{#1}}}

\begin{table}[htp]
\begin{center}
\begin{tabular}{| l c |  c || r | r || r | r |}
\hline
  & $\epsilon$-order&Dim& \thead{\vegas \\($\sigma = 10^{-3}$)}  &  \thead{\iflow \\($\sigma = 10^{-3}$)}& \thead{\vegas \\($\sigma = 10^{-4}$)}  &  \thead{\iflow \\($\sigma = 10^{-4}$)} \\ 
  \hline
  
  \multirow{3}{4em}{$\bK_{00;00111}$}
& -1& 2 & \nprintbf{135000} & \nprint{614400} & \nprint{2475000} & \nprintbf{1830912}
\\
& 0& 2 & \nprintbf{220000} & \nprint{819200} & \nprint{3510000} & \nprintbf{2314240}
\\
& 1& 2 & \nprintbf{270000} & \nprint{811008} & \nprint{6370000} & \nprintbf{2969600}
\\
\hline

\multirow{3}{4em}{$\bK_{00;11011}$}
& 0& 3 & \nprintbf{270000} & \nprint{778240} & \nprint{13135000} & \nprintbf{8036352}
\\
& 1& 3 & \nprintbf{325000} & \nprint{839680} & \nprint{18700000}0 & \nprintbf{8282112}
\\
& 2& 3 & \nprintbf{760000} & \nprint{937984} & \nprint{40635000} & \nprintbf{8740864}
\\
\hline

\multirow{3}{4em}{$\bK^{(\pm)}_{01;00111}$}
& -1& 2 & \nprintbf{135000} & \nprint{454656} & \nprint{3145000} & \nprintbf{1146880}
\\
& 0& 3 & \nprint{3895000} & \nprintbf{3641344} & \nprint{363850000} & \nprintbf{279408640}
\\
& 1& 3 & \nprint{30520000} & \nprintbf{26243072} & - &-
\\
\hline

\multirow{3}{4em}{$\bK^{(\pm)}_{01;11011}$}
& -1& 3 & \nprintbf{450000} & \nprint{757760} & \nprint{36900000} & \nprintbf{24240128}
\\
& 0& 4 & \nprint{13870000} & \nprintbf{11059200} & \nprint{1312245000} & \nprintbf{946786304}
\\
& 1& 4 & \nprint{9145000} & \nprintbf{7147520} & \nprint{865825000} & \nprintbf{172482560}
\\
\hline

\multirow{3}{4em}{$\bK^{(\pm)}_{01;10110}$}
& -1& 2 & \nprintbf{70000} & \nprint{208896} & \nprint{2475000} & \nprintbf{1019904}
\\
& 0& 3 & \nprintbf{220000} & \nprint{450560} & \nprint{12420000} & \nprintbf{2867200}
\\
& 1& 3 & \nprintbf{385000} & \nprint{528384} & \nprint{28350000} & \nprintbf{2887680}
\\
\hline

\multirow{3}{4em}{$\bK^{(+-)}_{11;00111}$}
& -2& 2 & \nprintbf{70000} & \nprint{245760} & \nprint{1885000} & \nprintbf{1130496}
\\
& -1& 4 & \nprintbf{1150000} & \nprint{1306624}  & \nprint{108675000} & \nprintbf{83521536}
\\
& 0& 4 & \nprint{125995000} & \nprintbf{102195200} & - &-
\\
\hline

\multirow{3}{4em}{$\bK^{(++)}_{11;00111}$}
& -2& 2 & \nprintbf{70000} & \nprint{196608} & \nprint{1375000} &\nprintbf{1011712}
\\
& -1& 4 & \nprintbf{450000} & \nprint{536576} & \nprint{37510000} & \nprintbf{24129536}
\\
& 0& 4 & \nprint{38745000} & \nprintbf{35098624} & - &-
\\
\hline

\multirow{3}{4em}{$\bK^{(++)}_{11;11011}$}
& -2& 3 & \nprintbf{135000} & \nprint{249856} & \nprint{11385000} & \nprintbf{10633216}
\\
& -1& 5 & \nprint{1150000} & \nprintbf{1138688} & \nprint{115020000} & \nprintbf{93896704}
\\
& 0& 5 & 8260000 & \nprintbf{7741440} & \nprint{802300000} & \nprintbf{713129984}
\\
\hline

\multirow{3}{4em}{$\bK^{(\pm)}_{02;10110}$}
& -1& 2 & \nprintbf{100000} & \nprint{385024} & \nprint{3145000} & \nprintbf{1048576}
\\
& 0& 3 & \nprintbf{850000} & \nprint{1085440} & \nprint{76995000} & \nprintbf{61423616}
\\
& 1& 3 & \nprint{5400000} & \nprintbf{5062656} & \nprint{505120000} & \nprintbf{388235264}
\\
\hline

\end{tabular}
\end{center}
\caption{We list the number of integrand evaluations needed to reach $10^{-3}$ and $10^{-4}$ precision for the two-loop integrals using \vegas and \iflow. 
The first column shows the order in $\epsilon$ after sector decomposition (see Sec.~\ref{sec:theory}) and the second column the integral dimensionality in parametrized form. 
Empty entries correspond to integrals we ignored since \iflow runs into memory problems, as reported in \cite{Gao:2020zvv}. }
\label{tab:results2L}
\end{table}

\begin{table}[htp]
\begin{center}
\begin{tabular}{| l c |  c || r | r || r | r |}
\hline
& $\epsilon$-order&Dim &\thead{\vegas \\($\sigma = 10^{-3}$)}  &  \thead{\iflow \\($\sigma = 10^{-3}$)}& \thead{\vegas \\($\sigma = 10^{-4}$)}  &  \thead{\iflow \\($\sigma = 10^{-4}$)}
\\ 
\hline

\multirow{3}{2.1em}{$\vecbf{B}_0$}%{$K^{???}_{2;000;001100110}$}
& 0& 3 & \nprintbf{175000} & \nprint{659456} & \nprint{3895000} & \nprintbf{1507328}
\\
& 1& 3 & \nprintbf{220000} & \nprint{782336} & \nprint{5635000} & \nprintbf{2072576}
\\
& 2& 3 & \nprintbf{325000} & \nprint{888832}  & \nprint{8260000} & \nprintbf{2625536}
\\
\hline

\multirow{3}{2.1em}{$\vecbf{B}_1$}%{ $K^{??\pm}_{2;001;001100110}$}
& -2& 2 & \nprintbf{135000} & \nprint{610304} & \nprint{2320000} & \nprintbf{1409024}
\\
& -1& 4 & \nprintbf{270000} & \nprint{602112} & \nprint{11725000} & \nprintbf{2445312}
\\
& 0& 4 & \nprintbf{760000} & \nprint{1024000} & \nprint{51475000} & \nprintbf{32100352}
\\
\hline

\multirow{3}{2.1em}{$\vecbf{B}_2$}%{ $K^{??\pm}_{1;001;010100011}$}
& -1& 3 & \nprintbf{175000} & \nprint{487424} & \nprint{5635000} & \nprintbf{1536000}
\\
& 0& 4 & \nprintbf{270000} & \nprint{655360} & \nprint{11385000} & \nprintbf{2076672}
\\
& 1& 4 & \nprintbf{385000} & \nprint{667648} & \nprint{16195000} & \nprintbf{2539520}
\\
\hline

\multirow{3}{2.1em}{$\vecbf{B}_3^+$}%{$K^{?++}_{2;011;001110001}$}
& -2& 3 & \nprintbf{135000} & \nprint{442368} & \nprint{4300000} & \nprintbf{2441216}
\\
& -1& 5 & \nprintbf{1750000} & \nprint{1777664} & \nprint{165760000} & \nprintbf{118611968}
\\
& 0& 5 &  \nprint{4945000} & \nprintbf{4096000} & \nprintbf{47197000} & \nprint{308641792}
\\
\hline

\multirow{3}{2.1em}{$\vecbf{B}_3^-$}%{$K^{?+-}_{2;011;001110001}$}
& -2& 3 & \nprintbf{175000} & \nprint{528384} & \nprint{4300000} & \nprintbf{2146304}
\\
& -1& 5 & \nprintbf{1620000} & \nprint{1757184}  & \nprint{154375000} & \nprintbf{112689152}
\\
& 0& 5 & - & - & - &-
\\
\hline

\multirow{3}{2.1em}{$\vecbf{B}_4^+$}%{$K^{?++}_{1;011;001100110}$}
& -2& 3 & \nprintbf{100000} & \nprint{405504} & \nprint{2800000} & \nprintbf{2142208}
\\
& -1& 5 & \nprintbf{595000} & \nprint{1007616} & \nprintbf{47950000} & \nprint{51929088}
\\
& 0& 5 & \nprintbf{4300000} & \nprint{4689920} & \nprint{425385000} & \nprintbf{363270144}
\\
\hline

\multirow{3}{2em}{$\vecbf{B}_4^-$}%{$K^{?+-}_{1;011;001100110}$}
& -2& 3 & \nprintbf{135000} & \nprint{438272} & \nprint{3700000} & \nprintbf{2392064}
\\
& -1& 5 & \nprintbf{325000} & \nprint{569344} & \nprint{26775000} & \nprintbf{16392192}
\\
& 0& 5 & \nprint{32200000} & \nprintbf{28790784} & - &-
\\
\hline

\multirow{3}{2.1em}{$\vecbf{B}_5^{++}$}%{$K^{+++}_{1;111;001010101}$}
& -3& 3 & \nprintbf{100000} & \nprint{376832} & \nprint{4725000} &  \nprintbf{1892352}
\\
& -2& 6 & \nprintbf{1495000} & \nprint{1650688} & \nprint{141010000} &  \nprintbf{115605504}
\\
& -1& 6 & \nprint{59670000} & \nprintbf{49348608} & - & -
\\
\hline

\multirow{3}{2.1em}{$\vecbf{C}_0$}%{$K^{???}_{2;000;001101011}$}
& -1& 3 & \nprintbf{220000} & \nprint{626688} & \nprint{5875000} & \nprintbf{2322432}
\\ 
& 0& 4 & \nprintbf{325000} & \nprint{774144} & \nprint{14625000} & \nprintbf{5808128}
\\
& 1& 4 & \nprintbf{595000} & \nprint{831488} & \nprint{26775000} & \nprintbf{8294400}
\\
\hline
  
\multirow{3}{2.1em}{$\vecbf{D}_0$}%{ $K^{???}_{1;000;001110011}$}
& 0& 4 & \nprintbf{270000} & \nprint{684032} & \nprint{10395000} & \nprintbf{4870144}
\\
& 1& 4 & \nprintbf{385000} & \nprint{790528} & \nprint{14245000} & \nprintbf{4898816}
\\
& 2& 4 & \nprintbf{595000} & \nprint{905216} & \nprint{23760000} & \nprintbf{5582848}
\\
\hline

\multirow{3}{2.1em}{$\vecbf{D}_1$}%{ $K^{??\pm}_{1;001;001110011}$}
& -1& 4 & \nprintbf{520000} & \nprint{827392} & \nprint{39370000} & \nprintbf{28872704}
\\
& 0& 5 & \nprint{5170000} & \nprintbf{4710400} & \nprint{485095000} & \nprintbf{331739136}
\\
& 1& 5 & \nprint{7975000} & \nprintbf{6582272} & \nprint{714220000} & \nprintbf{463904768}
\\
\hline

\end{tabular}
\end{center}
\caption{Results for the three-loop integrals. Same notation as in Table~\ref{tab:results2L}.}
\label{tab:results3L}
\end{table}

\begin{table}[ht]
\begin{center}
\begin{tabular}{ | l c |  c || r | r || r | r | }
\hline
& $\epsilon$-order&Dim &\thead{\vegas \\($\sigma = 10^{-3}$)}  &  \thead{\iflow \\($\sigma = 10^{-3}$)}& \thead{\vegas \\($\sigma = 10^{-4}$)}  &  \thead{\iflow \\($\sigma = 10^{-4}$)}
\\  
\hline
  
\multirow{3}{2.9em}{$\vecbf{M}_0$}%{ $K^{\pm\pm\pm\pm}_{1;0000;11111}$}
& -1& 4 & \nprintbf{220000} & \nprint{839680} & \nprint{5875000} & \nprintbf{2473984}
\\
&0& 4 & \nprintbf{325000} & \nprint{741376} & \nprint{7695000} & \nprintbf{2252800}
\\
& 1& 4 & \nprintbf{385000} & \nprint{970752} & \nprint{10075000} & \nprintbf{2813952}
\\
\hline

\multirow{3}{2.9em}{$\vecbf{M}_1$}%{ $K^{\pm\pm\pm\pm}_{1;1000;11111}$}
& -1 & 5 & \nprintbf{4725000} & \nprint{5513216} & \nprintbf{467635000} & \nprint{469925888}
\\
& 0 & 5 & \nprintbf{3700000} & \nprint{4268032} & \nprint{358150000} & \nprintbf{348610560}
\\
& 1 & 5 & \nprintbf{2170000} & \nprint{2498560} & \nprint{203770000} & \nprintbf{176631808}
\\
\hline  

\multirow{3}{2.9em}{$\vecbf{M}_2^{-}$}%{ $K^{+-++}_{1;1100;11111}$}
& -3& 3 & \nprintbf{175000} & \nprint{557056} & \nprint{4095000} & \nprintbf{1503232}
\\
& -2& 6 & \nprint{2320000} & \nprintbf{2105344} & \nprint{213885000} & \nprintbf{132751360}
\\
& -1& 6 & \nprint{119350000} & \nprintbf{96231424} & - & -
\\
\hline

\multirow{3}{2.9em}{$\vecbf{M}_2^{+}$}%{ $K^{+-++}_{1;1100;11111}$}
& -3& 3 & \nprintbf{175000} & \nprint{581632} & \nprint{4095000} & \nprintbf{1839104}
\\
& -2& 6 & \nprint{2635000} & \nprintbf{2314240} & \nprint{248845000} & \nprintbf{151486464}
\\
& -1& 6 & \nprint{27295000} & \nprintbf{22687744} & - & -
\\
\hline

\multirow{3}{2.9em}{$\vecbf{M}_3^{+}$}%{ $K^{\pm\pm\pm\pm}_{1;1100;11111}$}
& -3& 3 & \nprintbf{175000} & \nprint{577536} & \nprint{4095000} & \nprintbf{1413120}
\\
& -2& 6 & \nprint{2970000} & \nprintbf{2588672} & \nprint{298420000} & \nprintbf{183275520}
\\
& -1& 6 & \nprint{28350000} & \nprintbf{24297472} & - & - 
\\
\hline

\multirow{3}{2.9em}{$\vecbf{M}_3^{-}$}%{ $K^{+-++}_{1;1100;11111}$}
& -3& 3 & \nprintbf{175000} & \nprint{561152} & \nprint{5170000} & \nprintbf{1470464}
\\
& -2& 6 & \nprintbf{1045000} & \nprint{1048576} & \nprint{86950000} & \nprintbf{44961792}
\\
& -1& 6 & \nprint{23760000} & \nprintbf{20635648} & - & - 
\\
\hline

\multirow{3}{2.9em}{$\vecbf{M}_4^{+++}$}%{ $K4_1_pppm_111111111_v2$}
& -4& 4 & \nprintbf{175000} & \nprint{471040} & \nprint{7420000} & \nprintbf{2490368}
\\
& -3& 8 & \nprint{1885000} & \nprintbf{1835008} & \nprint{181570000} & \nprintbf{115736576}
\\
& -2& 8 & \nprint{18270000} & \nprintbf{13864960} & - & -
\\
\hline
\end{tabular}
\end{center}
\caption{Results for the four-loop integrals. Same notation as in Table~\ref{tab:results2L}.}
\label{tab:results4L}
\end{table}

We continue in this subsection by showing explicit results for the comparison of \vegas and \iflow for the two-, three-, and four-loop integrals introduced in Sec.~\ref{sec:theory}.
For this comparison we present the number of integrand evaluations needed for \iflow\footnote{\iflow evaluations include the pre-training stage mentioned in Sec.~\ref{sec:setup}.} and \vegas for each integrand at each order in epsilon and relative precision $\sigma = 10^{-3}$ and $10^{-4}$. 
We note that comparing the computational time is not a satisfactory metric. 
\vegas has been substantially optimized and its performance is fully parallelized. 
Even though we have parallelized the \iflow sampling, there is still plenty of room to improve its performance on an implementation level.
Moreover, the training stage of \iflow is not the limiting part of the algorithm and sampling is by far the most time consuming part. 
Therefore, the sampling number is a more coherent metric, akin as done in previous comparisons \cite{Bendavid:2017zhk, Gao:2020zvv}.

The results are summarized in Tables~\ref{tab:results2L}, \ref{tab:results3L}, and \ref{tab:results4L} for the two-, three-, and four-loop integrals respectively. 
We note that we were not able to estimate some higher-order-in-$\epsilon$ terms, since \iflow computations lead to memory problems akin as reported by \cite{Gao:2020zvv}. For $10^{-3}$ relative precision, \vegas often required fewer evaluations compared to \iflow, especially for lower-dimensional integrals. 
When increasing the complexity of the integrand (either by increasing the loop order, the integral dimensionality, the $\epsilon$-order or by requiring more precision) \iflow starts to pass \vegas.
This is consistent with the observations presented in \cite{Gao:2020vdv}. 
When requiring $10^{-4}$ precision, \iflow has outperformed \vegas for almost all cases and orders in $\epsilon$.

\begin{figure}[ht]
\centering  
\includegraphics[width=0.41\textwidth]{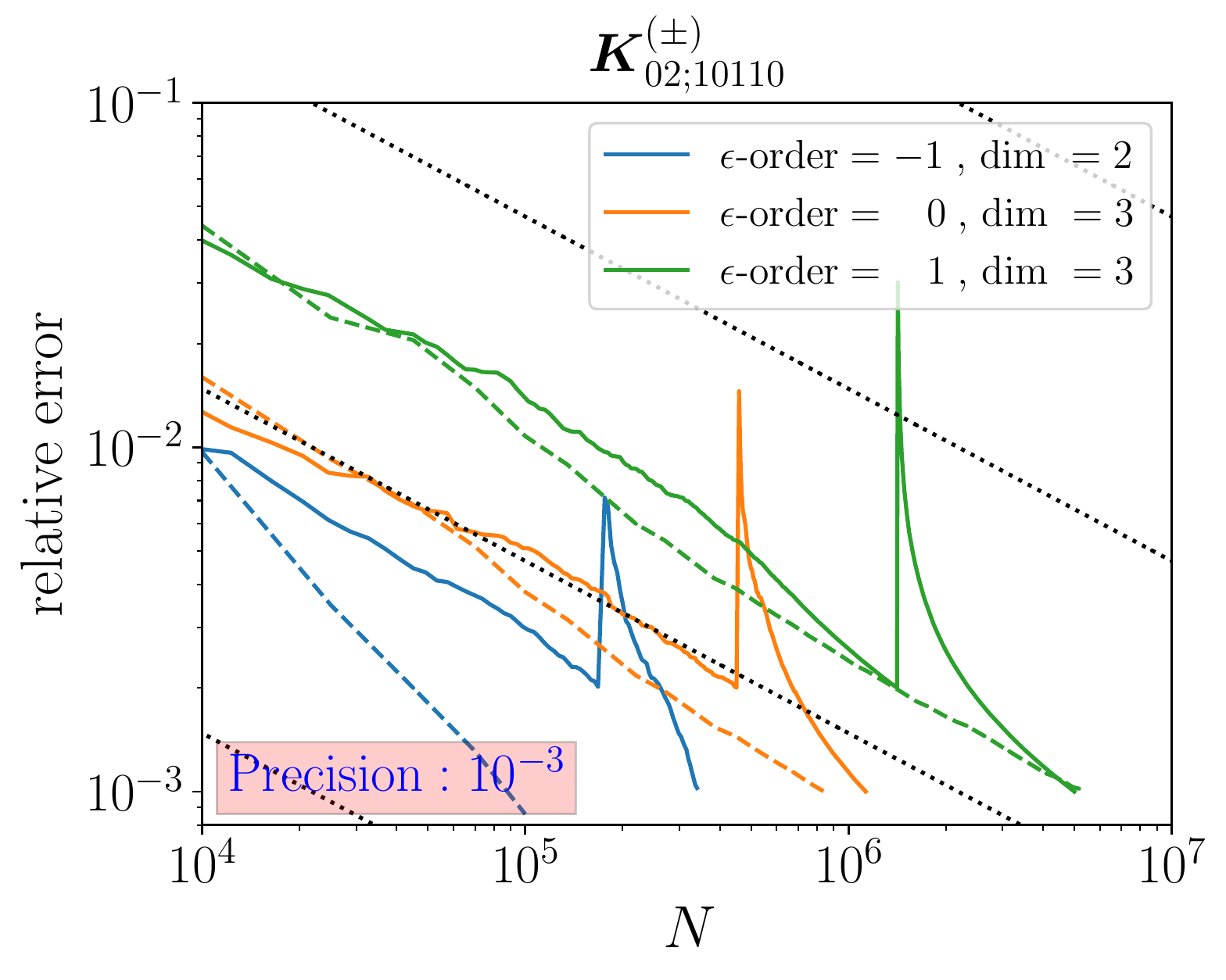}
\includegraphics[width=0.41\textwidth]{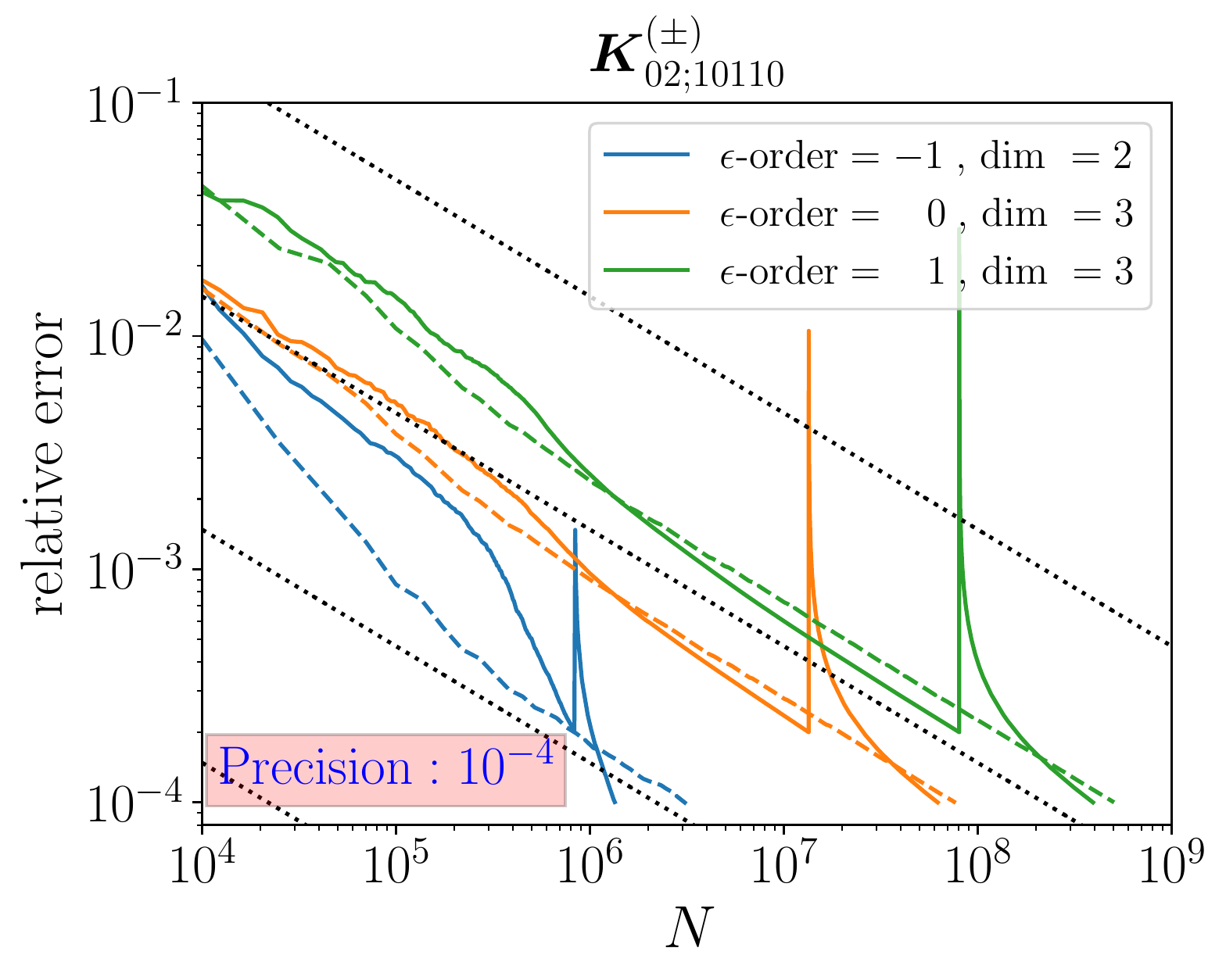}
\includegraphics[width=0.41\textwidth]{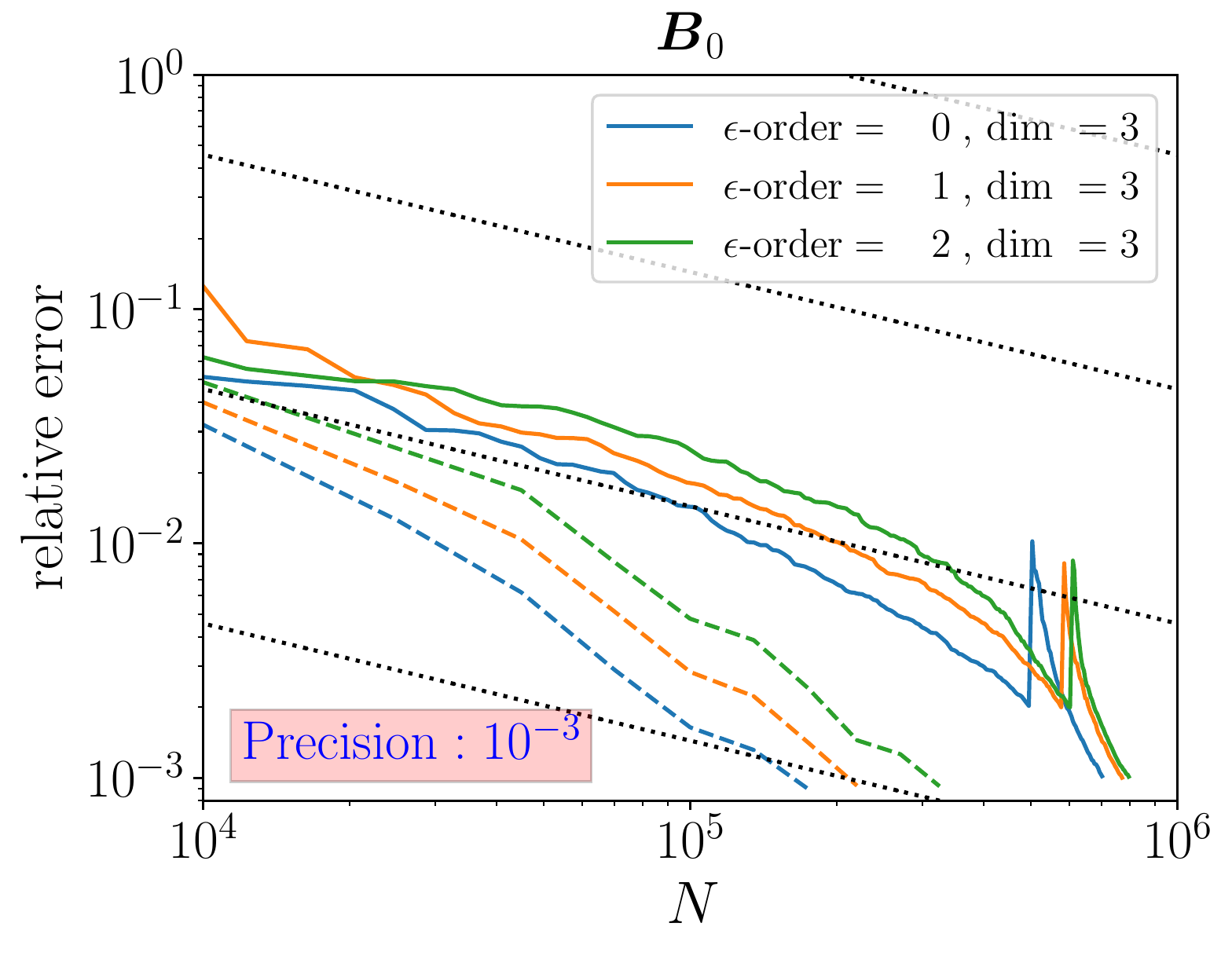}
\includegraphics[width=0.41\textwidth]{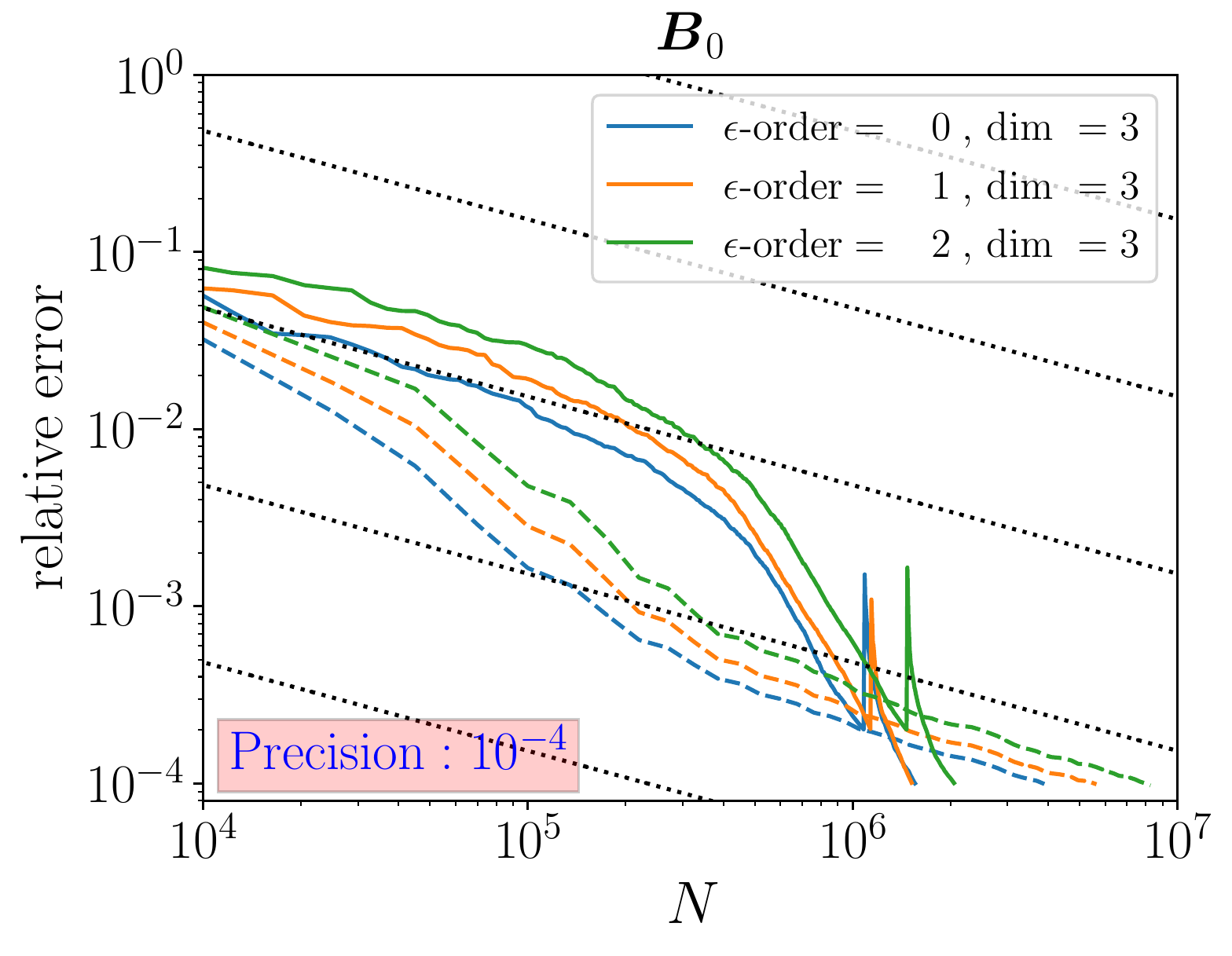}
\caption{\label{fig:errors_2l}
  \small This figure shows the evolution of the relative error with the number of iterations for \iflow (solid lines) and \vegas (dashed lines). 
  The two top figures correspond to the two-loop $\bK^{(\pm)}_{02;10110}$ integral as an illustration of a $G^3$-order integral and on the bottom we show the three-loop integral $B_0$ for an example at $G^4$. 
  The left plots are for a relative precision of $10^{-3}$ and the right panels for $10^{-4}$. Dotted lines indicate the theoretically expected $1/\sqrt{N}$ scaling once the integrator stops learning about the phase-space distribution and only samples more points. 
  The discontinuity in \iflow lines are due to a burn-in (pre-training) stage, see Sec.~\ref{sec:setup}.
}
\end{figure}

In order to understand the scaling behaviour of the relative error, we display in Fig.~\ref{fig:errors_2l} the evolution of the \iflow (solid lines) and \vegas (dashed lines) error as a function of the number of evaluations $N$. 
We use the exemplary integrals $\bK^{(\pm)}_{02;10110}$ and $B_0$ defined in Eqs.\,\eqref{eq:K2_9} and \eqref{eq:3L_B} respectively.
The discontinuities for the \iflow graphs are due to the pre-training stage. 
We observe that for both integrals at $10^{-3}$, \vegas indeed reaches the required precision faster than \iflow.
\iflow underperforms here due to the early stage of learning the phase-space distribution that already requires a high number of evaluations.
Also for $10^{-4}$ precision \iflow still has a latent training stage, but once it is fully trained the error graph is significantly steeper as compared to \vegas, especially for the harder three-loop integral. 
The dotted lines represent the expected $\sigma = 1/\sqrt{N}$ behaviour according to Eq.~\eqref{eq:error_scaling} in a late phase where only extra sampling is being performed. 
Hence, the asymptoptic behavior of \vegas typically follows this $1/\sqrt{N}$ behaviour.
Differently, NNs have an asymptotic behaviour better than $1/\sqrt{N}$ since they continue gathering information and learn about the system even in the late stage. 
The expectation that NNs work better for more complex integrals is confirmed by these plots in Fig.\,\ref{fig:errors_2l}. 
Note that the maximal dimensionality of the example integrals in Fig.\,\ref{fig:errors_2l} is 3. 
When increasing the dimension of the integrals (typically when going to higher loops) the crossing-point in which \iflow outperforms \vegas happens earlier (see Table~\ref{tab:results3L}). 
For instance, for $\vecbf{B}_2$, the required number of evaluations is five times smaller for \iflow when estimating it with $10^{-4}$ precision. 
Therefore, when computing high-loop integrals NN technologies like \iflow are leading to significant improvements.  

To get yet another impression on the scaling behaviour we show in Fig.~\ref{fig:sigma_evolve} the total number of evaluations needed as a function of the relative precision required for two (simpler) integrals where we were able to push to an even higher relative precision $10^{-5}$.
The left (right) panel display the results for the two(three)-loop $\bK_{00;00111}$ ($D_0$) integral for the leading order term in $\epsilon$.
We observe that for the the two-loop integral \vegas' scaling follows the $\sigma = 1/\sqrt{N}$ line.
Trying to achieve $\sigma < 10^{-6}$ precision demands $\mathcal{O}(10^{10})$ evaluations, incurring into the memory bounds. 
Opposed to this, \iflow presents a surprisingly good scaling following the $1/N^2$ line.

For three loops the behaviour of \vegas is similar. 
\iflow, though, has shown here a similarly bad behaviour as \vegas starting from a required precision $\sigma < 10^{-4}$. 
However, we notice a slightly smaller slope indicating that the neural network still keeps learning about the phase-space. 
While one could claim that this indicates a saturation of the benefits of using neural networks, we stress that the NN architecture is the same for all integrals (it only changes according to the number of dimensions of the integral \cite{Gao:2020vdv}). 
Playing with the architecture may improve the training --- a more-in-depth analysis of the optimal strategy will be the focus of a future study. 
On the other side, we do not see any potential improvements that could be done for the \vegas setup that could substantially change its asymptotic scaling.

\begin{figure}[ht]
\centering  
\includegraphics[width=0.49\textwidth]{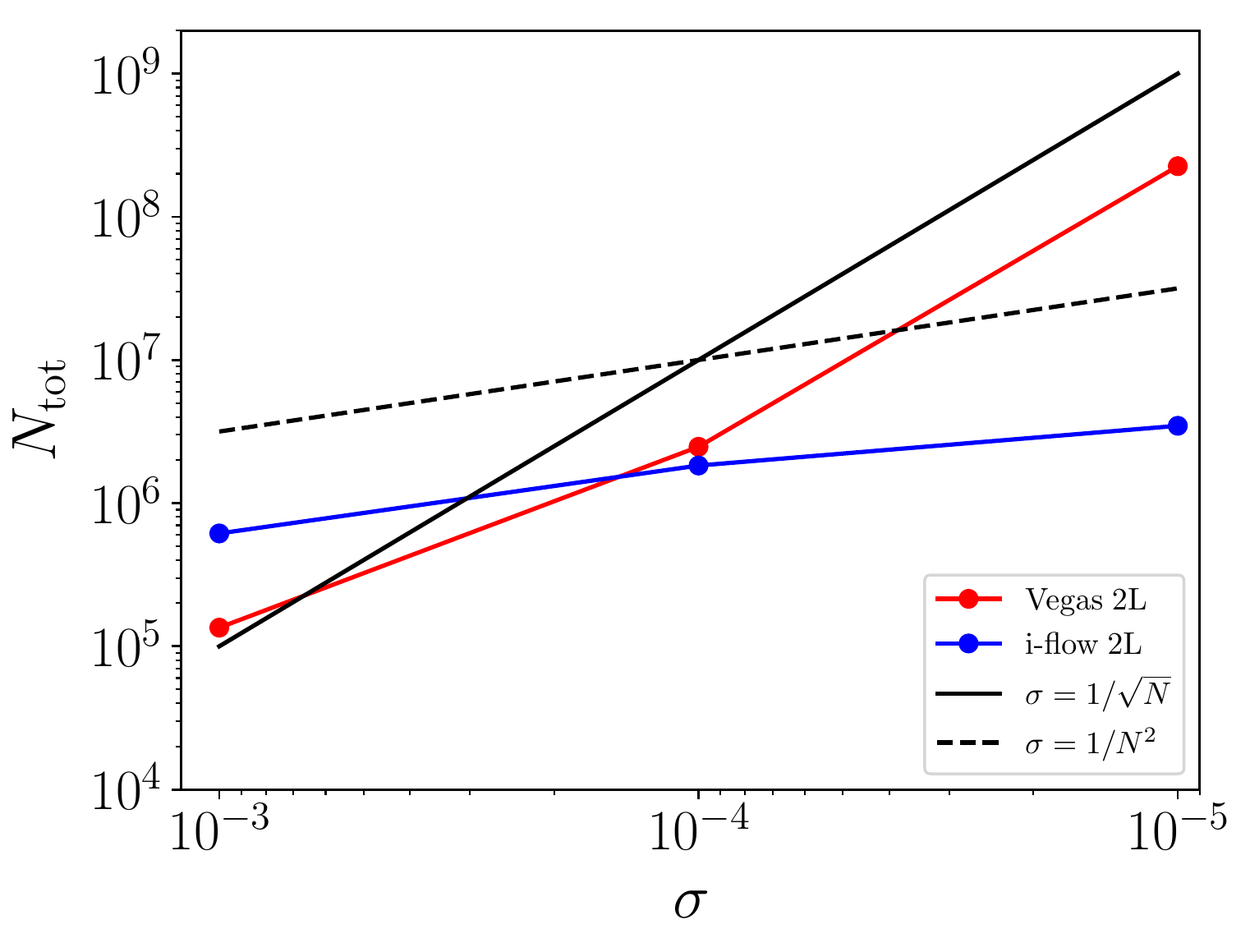}
\includegraphics[width=0.49\textwidth]{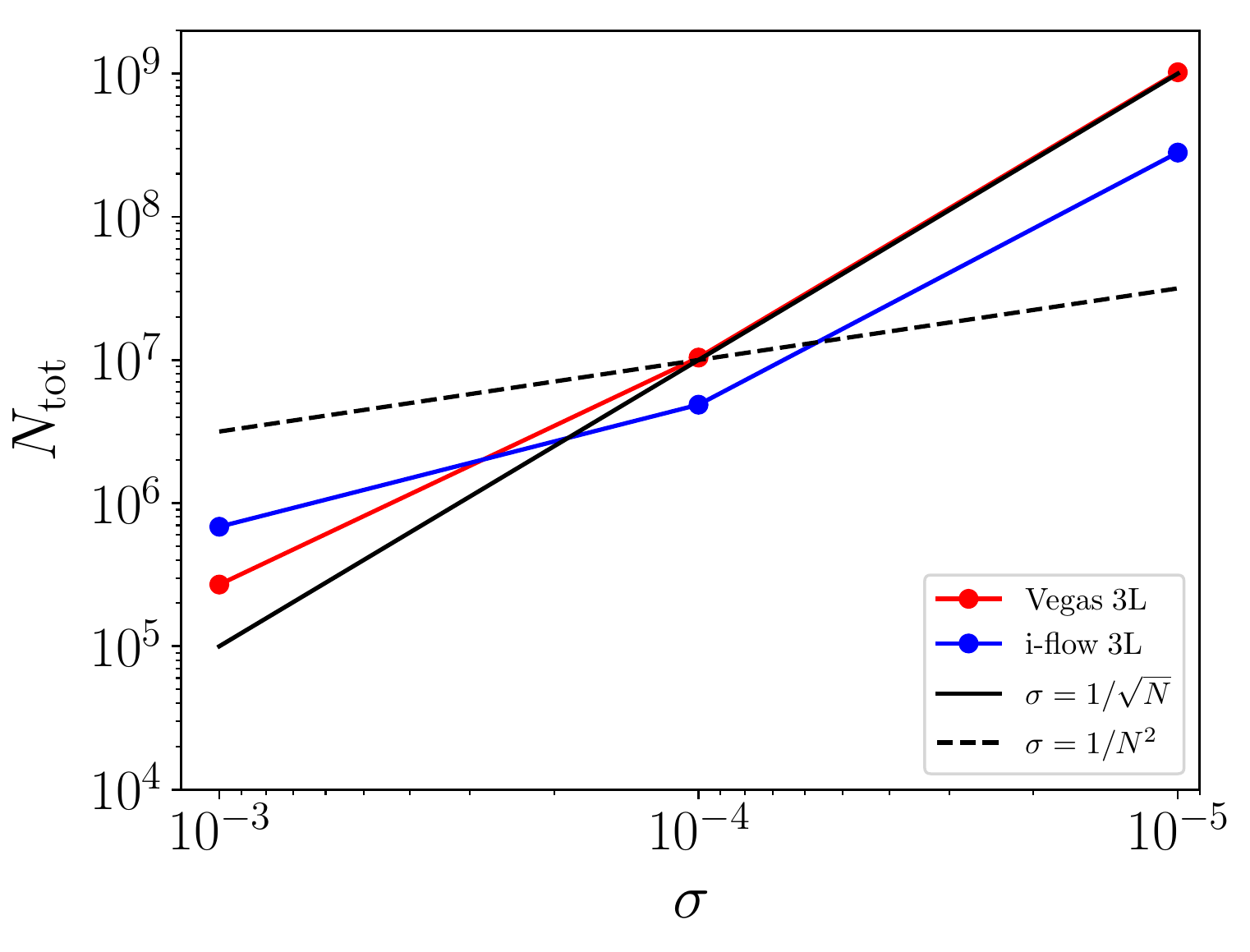}
\caption{\label{fig:sigma_evolve}
\small We plot the total number of evaluations $N_{\rm tot}$ as a function of the relative precision $\sigma$: 
On the left for the leading $\epsilon$ term of the two-loop integral $\bK_{00;00111}$; On the right for the leading term of the three-loop integral $D_0$. 
The black lines show expected theoretical behaviour for comparison, see main text.}
\end{figure}

\section{Discussion and Outlook}\label{sec:discussion}

In this work we have initiated the application of modern machine learning techniques to the numerical evaluation of multi-loop Feynman integrals, with a special focus on loop integrals relevant to make precision predictions for gravitational-wave observations.
Using \pysecdec's C++ interface for the sector decomposition and contour deformation we have compared two different Monte-Carlo integrators: the traditional \vegas method, based on partitioning the phase-space into non-uniform histograms and \iflow, a neural-network sampler that learns autonomously about the phase-space distribution of the integrand.
We want to emphasize that numerical approximations can be useful not only to check analytical expressions but also open up the stage for the use of high-precision numerical results in direct numerical construction of gravitation waveform templates or integer relation conjectures for analytical reconstruction.

We have found that for simpler integrals, namely lower-dimensional, lower order in $\epsilon$, and integrals containing fewer linear propagators, \vegas performs better.
This is partially due to a learning phase that is required for an unbiased neural network setup.
However, increasing the complexity of the phase-space or aiming to surpass per mille precision makes integration with \vegas significantly more time-consuming.
\iflow starts in such cases to outperform traditional methods.
Based on normalizing-flows, \iflow provides an efficient and systematic method to sample the phase-space.
Our results are consistent with the previous observations of \iflow applied to other systems: its error scales slowly in the early stages due to an initial transient phase, but the normalizing flow keeps learning about the integrand topology.
Due to its sampling strategy, \iflow's variance estimate then generically decreases faster than the naive $\frac{1}{\sqrt{N}}$ for traditional Monte-Carlo sampling, where $N$ is the number of integrand evaluations.

We would like to point out the current limitations for numerical integration via sector decomposition and Monte-Carlo methods:
First, our sector decomposed integrands tend to run into divergences (undetected singularities) that need to be taken care of manually.
Second, requiring more precision ($\sigma < 10^{-4}$) demands $\mathcal{O}(10^{10})$ evaluations meaning that we hit a hardware wall in terms of memory requirement for \iflow.
Improved sector decomposition algorithms have the potential to not only overcome the former, but can lead to better integrands when it comes to convergence speed, which in turn reduces the number of required integrand evaluations. 
The memory issues of \iflow can be fixed with an improved memory management, i.e. only storing results where required.

One idea of improvement of our current setup for PM integrals is to utilize integral identities like Eq.~\eqref{eq:B3B4id} to identify a set of independent integrals that have desirable properties for numerical algorithms.
Of course, this could simply be done by trial-and-error, but it would also be interesting to have integrand-level criteria to determine whether a given integral is suited for numerical integration or not.
One trivial criterion that we have identified is the positiveness of the Symanzik polynomials of the parametrized form of the integral.
Positive polynomials render complex contour deformations unnecessary and can significantly decrease the integrand evaluation time and improve its convergence properties.
We have analytically continued the external kinematics in order to achieve a positive $\cF$ polynomial for many of our examples.

As a further improvement, we note that \pysecdec has recently been extended by a quasi-Monte Carlo (QMC)~\cite{MOROKOFF1995218} integrator~\cite{Borowka:2018goh}.
QMC uses quasi-random grids to generate sample points in the phase space, while traditional MC samples random numbers.
This improves the theoretical scaling of the variance from $\frac{1}{\sqrt{N}}$ for traditional Monte-Carlo to $\frac{1}{N}$ or even $\frac{1}{N^2}$.
A challenge for QMC algorithms is the exponential scaling of the variance in the integral dimension $d$.
For the foreseeable future we do not expect to find integrals with dimension significantly higher than 10, for which methods have been developed to overcome this scaling~\cite{Borowka:2018goh}.
We hence expect that a QMC integrator could be combined with improved NN phase-space sampling to reach an even better performance (see e.g. \cite{QMCflow}).
We leave that for future work.

It is clear that the framework developed in this paper is straightforwardly applicable to other multi-loop integrals, e.g.~in the context of the effective field theory of large-scale structure \cite{Baumann:2010tm, Carrasco:2012cv, Carrasco:2013mua, Konstandin:2019bay,Rubira:2020inb,Mergulhao:2021kip}. 
The success of similar methods for similar integration problems \cite{Bendavid:2017zhk, Klimek:2018mza, Chen:2020nfb, Bothmann:2020ywa} and in other areas, such as precise measurements at high-energy colliders \cite{Bishara:2019iwh, Gao:2020zvv, Otten:2019hhl, Danziger:2021eeg, DiSipio:2019imz, Butter:2019cae}, strongly motivate us to apply \it normalizing flows \rm to extending the multi-loop program in the context of gravitational waves. 
This work intends to be a beginning of an agenda in which numerical calculations and analytical results are complementary and together push forward the theory to exquisite precision.

%%%%%%%%%%%%%%%%%%%%%%%%%%%%%%%%%%%%%%%%%%%%%%%%%%
\section*{Acknowledgment}
%%%%%%%%%%%%%%%%%%%%%%%%%%%%%%%%%%%%%%%%%%%%%%%%%%

The authors are grateful to Stephen Jones, Go Mishima, Andres P\"oldaro, Vladyslav Shta\-bo\-ven\-ko for helpful correspondence on \texttt{pySecDec},
and to Joshua Isaacson for the support with iflow.
We are indebted to Luisa Lucie-Smith for the very effective comments on the draft.
We thank Christoph Dlapa and Rafael Porto for useful discussions and collaborations on related topics.
This work was supported by the Deutsche Forschungsgemeinschaft under Germany's Excellence Strategy EXC 2121 `Quantum Universe' (No.\,390833306) and EXC 2094 `ORIGINS' (No.\,390783311).
The work of RJ is supported by the grants IFT Centro de Excelencia Severo Ochoa SEV-2016-0597, CEX2020-001007-S and by PID2019-110058GB-C22 funded by MCIN/AEI/10.13039/501100011033 and by ERDF.
The work of RJ is supported by Grants-in-Aid for JSPS Overseas Research Fellow (No.\,201960698).
GK received support from the ERC-CoG Precision Gravity: From the LHC to LISA provided by the European Research Council (ERC) under the European Union’s H2020 research and innovation programme (grant No.\,817791).

\appendix
\section{Analytic derivations}\label{sec:appA}

In this appendix, we provide derivations for the analytic expression for two special classes of integrals.
We first consider $n$-loop integrals with $n{+}1$ massless squared propagators that form a {\it banana} topology and $n$ linear propagators of form $\vecbf{\ell}_{i\cdots j}\cdot \vecbf{u}$ (with $\vecbf{u}$ the unit vector in the $z$-direction).
In the second subsection we compute $n$-loop massless {\it banana} integrals with exactly two linear propagators. 

\subsection{Some symmetrization magic}
In this subsection, we present a unified framework to derive analytic expressions for $n$-loop banana integrals with $n$ linear propagators, including the one-loop $\vecbf{A}_{111}$ \eqref{pm-1-loop}, the two-loop integrals in \eqref{eq:K2_6} and \eqref{eq:K2_7}, the three-loop integrals $\vecbf{B}_5$ and $\vecbf{B}_6$, and the four-loop $\vecbf{M}_4$ in Section \ref{sec:theory}.

The key idea is to introduce an auxiliary loop integration that is fully localized by a $d$-dimensional $\delta$-distribution such that we can write the squared-propagator part of the integrand in a fully symmetric form in all loop momenta (including the auxiliary loop variable), i.e.
\begin{align}\label{}
{1 \over \vecbf{\ell}_1^2 \vecbf{\ell}_2^2 \cdots \vecbf{\ell}_n^2\, (\vecbf{\ell}_{1\cdots n} {-} \vecbf{q})^2}
= \int \dd^d\ell_{n+1}\,
{\delta^{(d)}(\vecbf{\ell}_{n+1} + \vecbf{\ell}_{1\cdots n} {-} \vecbf{q}) \over \vecbf{\ell}_1^2 \vecbf{\ell}_2^2 \cdots \vecbf{\ell}_n^2 \vecbf{\ell}_{n+1}^2}\,.
\end{align}
The resulting integral is invariant under the permutation of loop momenta; and thus we can write the integral as a full symmetric form in all loop momenta.
The essential observation is that we may write the sum of all permutations of the linear propagators as a product of Dirac-$\delta$ functions of the form $\delta(\ell_1^z)\cdots \delta(\ell_{n+1}^z)$.
As a result, all $\ell_i^z$ integrals can be resolved via these Dirac-$\delta$ functions, and the integral gets reduced to an ordinary massless banana integral in $d{-}1$ dimensions.

To illustrate our idea explicitly, let us work through the one-loop case:
\begin{align}\label{}
\int {\dd^{d} \ell \over \pi^{d/2}}\, {1 \over (\ell^z {-} i0)\, \vecbf{\ell}^2\, (\vecbf{\ell} {-} \vecbf{q})^2}
&= {1 \over 2}\int {\dd^{d}\ell_1 \dd^{d}\ell_2 \over \pi^{d/2}}
\bigg({1 \over \ell_1^z {-} i0} + {1 \over \ell_2^z {-} i0}\bigg)\,
{\delta^{(d)}(\vecbf{\ell}_1 {+} \vecbf{\ell}_2 {-} \vecbf{q}) \over \vecbf{\ell}_1^2\, \vecbf{\ell}_2^2}
\\[0.35em]
&= {2\pi i \over 2}\int {\dd^{d}\ell_1 \dd^{d}\ell_2  \over \pi^{d/2}}
{\delta(\ell_1^z)\delta(\ell_2^z)\,\delta^{(d)}(\vecbf{\ell}_1 {+} \vecbf{\ell}_2 {-} \vecbf{q}) \over \vecbf{\ell}_1^2\, \vecbf{\ell}_2^2}
\nn\\[0.35em]
&= {2\pi i \over 2}\int {\dd^{d-1}\ell^\perp \over \pi^{d/2}}\,
{1 \over (\vecbf{\ell}^\perp)^2\,(\vecbf{\ell}^\perp {-} \vecbf{q})^2}
\nn\\[0.35em]
&= 
{i \sqrt{\pi}\, \Gamma^2(-\epsilon)\, \Gamma(\epsilon+1)  \over  \Gamma (-2\epsilon )}\,,\nn
\end{align}
where we used $\vecbf{q}\cdot\vecbf{u} = q^z=0$ and the identity \cite{Cheng:1987ga}
\begin{align}\label{delta-id-1}
\delta(z_1 + z_2)\bigg({1 \over z_1 - i0} + {1 \over z_2 - i0}\bigg) = 2\pi i\, \delta(z_1)\delta(z_2)\,.
\end{align}
Here and in the rest of this appendix the numbers $z_i \in \mathbb{R}$.

Next, consider the two-loop integral, $\bK_{11;00111}^{+\pm}$.
We will need the following identities:
\begin{align}
\label{delta-id-2-1}
\delta(z_1 {+} z_2 {+} z_3) \left( {1 \over z_1 {-} i0}\frac{1}{z_{12} {-} i0} + \text{perms} \right) 
&= {(2\pi i)^2}\, \delta(z_1)\delta(z_2)\delta(z_3)\,,
\\[0.35 em]
\label{delta-id-2-2}
\delta(z_1 {+} z_2 {+} z_3) \left( {1 \over z_1 {-} i0}\frac{1}{z_{2} {-} i0} + \text{perms} \right) 
&= 2{(2\pi i)^2}\, \delta(z_1)\delta(z_2)\delta(z_3)\,,
\end{align}
where $z_{i\cdots j} = z_i + \cdots + z_j$ and ``perms'' denotes all permutations in all three variables $z_i$.
Following the procedure described above we easily arrive at
\begin{align}
\bK^{(++)}_{11;00111} = 2 \bK^{(+-)}_{11;00111}
&=  2\times {(2\pi i)^2 \over 6}  \int {\dd^{d-1} \ell^\perp_1 \dd^{d-1} \ell^\perp_2 \over \pi^d}
{e^{2 \gamma_E \epsilon} \over (\vecbf{\ell}_1^\perp)^2\, (\vecbf{\ell}_2^\perp)^2\, (\vecbf{\ell}_{12}^\perp {-} \vecbf{q})^2}
\\[0.35 em]
&=
- {4\pi\, e^{2 \gamma_E \epsilon}  \over 3}\,
{\Gamma^3(-\epsilon)\, \Gamma(1 + 2\epsilon) \over \Gamma(-3\epsilon )}.
\label{app-res-sunrise-2linear}
\end{align}
The derivation of identities \eqref{delta-id-2-1}, \eqref{delta-id-2-2} follows the method presented in Appendix A in \cite{Saotome:2012vy}.
More interestingly, using a similar method we find many identities of the type \eqref{delta-id-1}, \eqref{delta-id-2-1} and \eqref{delta-id-2-2}, which leads to an elegant derivation of many analytic results for similar integrals at higher-loop levels, e.g.\,$\vecbf{B}_5$, $\vecbf{B}_6$ and $\vecbf{M}_4$.
We list some identities of this type here:
\begingroup
\allowdisplaybreaks
\begin{align*}\label{}
{\delta(z_{1234}) \over (z_1 {-} i0) (z_{12} {-} i0) (-z_4 {-} i0)} + \text{perms}
&= {(2\pi i)^{3}} \delta(z_1)\delta(z_2)\delta(z_3)\delta(z_4)\,,
\\[0.3 em]
{\delta(z_{1234}) \over (z_1 {-} i0) (z_{12} {-} i0) (z_4 {-} i0)} + \text{perms}
&= 3\, (2\pi i)^{3} \delta(z_1)\delta(z_2)\delta(z_3)\delta(z_4)\,,
\\[0.3 em]
{\delta(z_{1234}) \over (z_1 {-} i0) (-z_{12} {-} i0) (-z_4 {-} i0)} + \text{perms}
&= 5\, (2\pi i)^{3} \delta(z_1)\delta(z_2)\delta(z_3)\delta(z_4)\,,
\\[0.3 em]
{\delta(z_{1234}) \over (z_1 {-} i0) (-z_{12} {-} i0) (z_4 {-} i0)} + \text{perms}
&= 3\, (2\pi i)^{3} \delta(z_1)\delta(z_2)\delta(z_3)\delta(z_4)\,,
\\[0.35 em]
%%%
%%%
{\delta(z_{1234}) \over (z_1 - i0) (z_2 - i0) (z_3 - i0)} + \text{perms}
&= 6\, (2\pi i)^{3} \delta(z_1)\delta(z_2)\delta(z_3)\delta(z_4)\,,
\\[0.3 em]
{\delta(z_{1234}) \over (z_1 - i0) (z_2 - i0) (-z_3 - i0)} + \text{perms}
&= 2\, (2\pi i)^{3} \delta(z_1)\delta(z_2)\delta(z_3)\delta(z_4)\,,
\\[0.3 em]
{\delta(z_{1234}) \over (z_1 - i0) (-z_2 - i0) (z_3 - i0)} + \text{perms}
&= 2\, (2\pi i)^{3} \delta(z_1)\delta(z_2)\delta(z_3)\delta(z_4)\,,
\\[0.3 em]
{\delta(z_{1234}) \over (z_1 - i0) (-z_2 - i0) (-z_3 - i0)} + \text{perms}
&= 2\, (2\pi i)^{3} \delta(z_1)\delta(z_2)\delta(z_3)\delta(z_4)\,,
\\[0.35 em]
%%%%%
%%%%%
%%%%%
{\delta(z_{12345}) \over (z_1 {-} i0) (z_{12} {-} i0) (z_{123} {-} i0) (-z_{5} {-} i0)} + \text{perms}
&= {(2\pi i)^{4}} \delta(z_1)\delta(z_2)\delta(z_3)\delta(z_4)\delta(z_5)\,,
\\[0.3 em]
{\delta(z_{12345}) \over (z_1 {-} i0) (z_{12} {-} i0) (z_{123} {-} i0) (z_{5} {-} i0)} + \text{perms}
&= 4\, {(2\pi i)^{4}} \delta(z_1)\delta(z_2)\delta(z_3)\delta(z_4)\delta(z_5)\,,
\\[0.3 em]
{\delta(z_{12345}) \over (z_1 {-} i0) (z_{12} {-} i0) (-z_{123} {-} i0) (-z_{5} {-} i0)} + \text{perms}
&= 9\, {(2\pi i)^{4}} \delta(z_1)\delta(z_2)\delta(z_3)\delta(z_4)\delta(z_5)\,,
\\[0.3 em]
{\delta(z_{12345}) \over (z_1 {-} i0) (z_{12} {-} i0) (-z_{123} {-} i0) (z_{5} {-} i0)} + \text{perms}
&= 6\, {(2\pi i)^{4}} \delta(z_1)\delta(z_2)\delta(z_3)\delta(z_4)\delta(z_5)\,,
\\[0.3 em]
{\delta(z_{12345}) \over (z_1 {-} i0) (-z_{12} {-} i0) (z_{123} {-} i0) (-z_{5} {-} i0)} + \text{perms}
&= 9\, {(2\pi i)^{4}} \delta(z_1)\delta(z_2)\delta(z_3)\delta(z_4)\delta(z_5)\,,
\\[0.3 em]
{\delta(z_{12345}) \over (z_1 {-} i0) (-z_{12} {-} i0) (z_{123} {-} i0) (z_{5} {-} i0)} + \text{perms}
&= 16\, {(2\pi i)^{4}} \delta(z_1)\delta(z_2)\delta(z_3)\delta(z_4)\delta(z_5)\,,
\\[0.3 em]
{\delta(z_{12345}) \over (z_1 {-} i0) (-z_{12} {-} i0) (-z_{123} {-} i0) (-z_{5} {-} i0)} + \text{perms}
&= 11\, {(2\pi i)^{4}} \delta(z_1)\delta(z_2)\delta(z_3)\delta(z_4)\delta(z_5)\,,
\\[0.3 em]
{\delta(z_{12345}) \over (z_1 {-} i0) (-z_{12} {-} i0) (-z_{123} {-} i0) (z_{5} {-} i0)} + \text{perms}
&= 4\, {(2\pi i)^{4}} \delta(z_1)\delta(z_2)\delta(z_3)\delta(z_4)\delta(z_5)\,.
\end{align*}
\endgroup

\vskip 10pt
\subsection{Some deformation magic}\label{app:deformation}

The goal of this subsection is to find analytic expressions for $\vecbf{B}_3$, $\vecbf{B}_4$, $\vecbf{M}_2$, $\vecbf{M}_3$ as well as $\bK_{11;00111}$ introduced in Section \ref{sec:theory}.
After integrating out up to two trivial loop momenta using the one-loop bubble formula, these can all be reduced to 
\begin{align}\label{}
&\bK^{(+\pm)}_{11;00111} = e^{2\epsilon\gamma_E}\,
\int{\dd^d\ell_1 \dd^d\ell_2 \over \pi^{d}}\,  {1 \over (\ell_1^z)\, (\pm\ell_2^z)}\,
{(\vecbf{q}^2)^{4 - d} \over \vecbf{\ell}_1^2\, \vecbf{\ell}_2^2\, (\vecbf{\ell}_{12} {-} \vecbf{q})^2}\,,
\\
&\vecbf{B}^{\pm}_{3} = e^{3\epsilon\gamma_E}\,
\frac{\Gamma^2({1/2} - \epsilon)\, \Gamma(1/2 + \epsilon)}{\Gamma(1-2 \epsilon)} \int{\dd^d\ell_1 \dd^d\ell_2 \over \pi^{d}}\,  {1 \over (\ell_1^z)\, (\pm\ell_2^z)}\,
{(\vecbf{q}^2)^{5 - 3d/2} \over \vecbf{\ell}_1^2\, \vecbf{\ell}_2^2\, [(\vecbf{\ell}_{12} {-} \vecbf{q})^2]^{(4-d)/2}}\,,
\\[0.5 em]
&\vecbf{B}^{\pm}_{4} = e^{3\epsilon\gamma_E}\,
\frac{\Gamma^2({1/2} - \epsilon)\, \Gamma(1/2 + \epsilon)}{\Gamma(1-2 \epsilon)} \int{\dd^d\ell_1 \dd^d\ell_2 \over \pi^{d}}\,  {1 \over (\ell_1^z)\, (\pm\ell_{12}^z)}\,
{(\vecbf{q}^2)^{5 - 3d/2} \over \vecbf{\ell}_1^2\, \vecbf{\ell}_2^2\, [(\vecbf{\ell}_{12} {-} \vecbf{q})^2]^{(4-d)/2}}\,,
\\[0.5 em]
&\vecbf{M}_2^\pm = e^{4\epsilon\gamma_E}\,
{\Gamma^3(1/2 - \epsilon)\, \Gamma(2\epsilon ) \over  \Gamma(3/2 - 3\epsilon)} \int {\dd^{d}\ell_1 \dd^{d}\ell_2 \over \pi^{d}}\, {1 \over (\ell_1^z) (\pm\ell_2^z)}\,
{(\vecbf{q}^2)^{6-2d} \over \vecbf{\ell}_1^2\, \vecbf{\ell}_2^2\, [(\vecbf{\ell}_{12} {-} \vecbf{q})^2]^{3-d}}\,,
\\[0.5 em]
&\vecbf{M}_3^\pm = e^{4\epsilon\gamma_E}\,
{\Gamma^3(1/2 - \epsilon)\, \Gamma(2\epsilon ) \over  \Gamma(3/2 - 3\epsilon)} \int {\dd^{d}\ell_1 \dd^{d}\ell_2 \over \pi^{d}}\, {1 \over (\ell_1^z) (\pm\ell_{12}^z)}\,
{(\vecbf{q}^2)^{6-2d} \over \vecbf{\ell}_1^2\, \vecbf{\ell}_2^2\, [(\vecbf{\ell}_{12} {-} \vecbf{q})^2]^{3-d}}\,.
\end{align}
Thus, it suffices to calculate
\begin{align}
\vecbf{I}_1^\pm
&\equiv
\int{\dd^d\ell_1 \dd^d\ell_2 \over \pi^{d}}\, {1 \over (\ell_1^z)\, (\pm\ell_2^z)}\, {(\vecbf{q}^2)^{3 - d + \nu} \over \vecbf{\ell}_1^2\, \vecbf{\ell}_2^2\, [(\vecbf{\ell}_{12} {-} \vecbf{q})^2]^\nu}\,,
\\
\vecbf{I}_2^\pm
&\equiv
\int{\dd^d\ell_1 \dd^d\ell_2 \over \pi^{d}}\, {1 \over (\ell_1^z)\, (\pm\ell_{12}^z)}\, {(\vecbf{q}^2)^{3 - d + \nu} \over \vecbf{\ell}_1^2\, \vecbf{\ell}_2^2\, [(\vecbf{\ell}_{12} {-} \vecbf{q})^2]^\nu}\,,
\end{align}
for $\nu=1$, $\nu=(4 {-} d)/2$, or $\nu = 3{-}d$.
While these integrals have a 2-loop massless banana topology plus two linear propagators, like $\bK_{11;00111}$, one of the three squared propagators has a non-integer power $\nu$, which we will leave generic in what follows.\footnote{
For a generic $\nu$ one has to be careful with possible analytic continuation throughout our derivations.}
Therefore, the method described in the previous subsection is not applicable because the integrands are no longer symmetric under permutation of the loop momenta.
We describe a method to perform a direct, analytical integration of the Feynman parametric representation\footnote{An alternative derivation that relies on the iterated-integration structure of hypergeometric functions is presented in \cite{Dlapa:2023hsl}.}.

Before moving on, we point out that $\vecbf{I}_1^+$ and $\vecbf{I}_2^+$ are related to each other whereas $\vecbf{I}_1^-$ and $\vecbf{I}_2^-$ are not.
To see this, a symmetrization of the first linear propagator leads to\footnote{This also provides an alternative proof for the relation $\vecbf{K}^{+-}_{11;00111} = {1\over 2}\vecbf{K}^{++}_{11;00111}$.}
\begin{align}
\vecbf{I}_2^\pm
&=
\frac{1}{2} \int{\dd^d\ell_1 \dd^d\ell_2 \over \pi^{d}}\, \left[ {1 \over (\ell_1^z - i 0)} + {1 \over (\ell_2^z - i 0)} \right] {1 \over (\pm\ell_{12}^z - i 0)}\, {(\vecbf{q}^2)^{3 - d + \nu} \over \vecbf{\ell}_1^2\, \vecbf{\ell}_2^2\, [(\vecbf{\ell}_{12} {-} \vecbf{q})^2]^\nu}
\nonumber \\
&=
\frac{1}{2} \int{\dd^d\ell_1 \dd^d\ell_2 \over \pi^{d}}\, {1 \over (\ell_1^z - i 0) (\ell_2^z - i 0)} {(\ell_{12}^z - i 0) \over (\pm\ell_{12}^z - i 0)}\, {(\vecbf{q}^2)^{3 - d + \nu} \over \vecbf{\ell}_1^2\, \vecbf{\ell}_2^2\, [(\vecbf{\ell}_{12} {-} \vecbf{q})^2]^\nu}\,.
\end{align}
For $\vecbf{I}_2^+$, the numerator cancels against the denominator, giving $\vecbf{I}_2^+ = \vecbf{I}_1^+ / 2$, while such cancellation does not occur for $\vecbf{I}_2^-$.
While often an $i 0$ in the numerator can be ignored it matters here: The integral above receives contributions mainly from ${\ell}_1^z \sim {\ell}_2^z \sim 0$ and behaves like $(0 - i 0) / (\pm 0 - i 0)$ in this region.
This is in a clear contrast to the case in which we have a structure like $({\rm finite} - i 0) / (0 - i 0)$ and the $i 0$ in the numerator can be safely neglected.

\paragraph{The first integral $\vecbf{I}_1^\pm$}

We denote the Feynman parameters by $x_1, \ldots , x_5$ corresponding to the five propagators.
The Feynman parametrization is then given by
\begin{align}
\vecbf{I}_1^\pm
&=
\frac{\Gamma_{4 - d + \nu}}{\Gamma_\nu}
\Bigg(\prod_{i=1}^{5} \int_0^\infty \dd x_i\Bigg)
%\int_0^\infty dx_2 \int_0^\infty dx_3 \int_0^\infty dx_4 \int_0^\infty dx_5~
\delta \left( 1 - \sum_{i \in I} x_i \right) x_5^{- 1 + \nu} \, \frac{{\cal U}^{4 - 3 d / 2 + \nu}}{({\cal F}^\pm)^{4 - d + \nu}}\,,
\end{align}
with Symanzik polynomials
\begin{align}
{\cal U}
&=
x_3 x_4 + x_4 x_5 + x_5 x_3\,,
\\
{\cal F}^\pm
&=
x_3 x_4 x_5 - \frac{1}{4} (x_1~x_2)
\left(
\begin{matrix}
x_4 + x_5 & \mp x_5
\\
\mp x_5 & x_3 + x_5
\end{matrix}
\right)
\left(
\begin{matrix}
x_1
\\
x_2
\end{matrix}
\right)
- i 0\,.
\end{align}
We do not yet specify the subset $I\subset\{1,2,\ldots,5\}$ since it can be chosen to be an arbitrary non-empty set according to the Cheng-Wu theorem \cite{Cheng:1987ga}.
For illustrative purposes we split the integral into two contributions from the two regions $x_3 > x_4$ and $x_3 < x_4$: $\vecbf{I}_1^\pm = \vecbf{I}_1^\pm \big|_{x_3 > x_4} + \vecbf{I}_1^\pm \big|_{x_3 < x_4}$.
Consider now $\vecbf{I}_1^+ \big|_{x_3 > x_4}$.
In order to understand how the integrations in $x_1$ and $x_2$ behave, we identify the directions that diagonalize the matrix
\begin{align}
\frac{1}{4} (x_1~x_2)
\left(
\begin{matrix}
x_4 + x_5 & \mp x_5
\\
\mp x_5 & x_3 + x_5
\end{matrix}
\right)
\left(
\begin{matrix}
x_1
\\
x_2
\end{matrix}
\right)
&=
x_3 x_4 x_5
(x_1'~x_2')
\left(
\begin{matrix}
\lambda_1 & 0
\\
0 & \lambda_2
\end{matrix}
\right)
\left(
\begin{matrix}
x_1'
\\
x_2'
\end{matrix}
\right),
\end{align}
with the rotation matrix
\begin{align}
\left(
\begin{matrix}
x_1'
\\
x_2'
\end{matrix}
\right)
&=
\left(
\begin{matrix}
\cos\theta & \sin\theta
\\
- \sin\theta & \cos\theta
\end{matrix}
\right)
\left(
\begin{matrix}
x_1
\\
x_2
\end{matrix}
\right),
~~~~
- \frac{\pi}{4} < \theta < \frac{\pi}{4}\,.
\end{align}
We further identify
\begin{align}
\sqrt{\lambda_1} x_1'
&\equiv
x_1''\,,
~~~~~~
\sqrt{\lambda_2} x_2'
\equiv
x_2''\,.
\end{align}
Note that $\lambda_2 > \lambda_1 > 0$ for $x_3 > x_4$.
Switching to $x_1''$ and $x_2''$ deforms the original integration region $[0,\infty)\times[0,\infty)$ for $x_1$ and $x_2$ in a nontrivial way, see Fig.\,\ref{fig:rotation}.
We call these regions $R^\pm$.
%%%%%%%%%%%%%%%%
\begin{figure}
\begin{center}
\includegraphics[width=0.8\columnwidth]{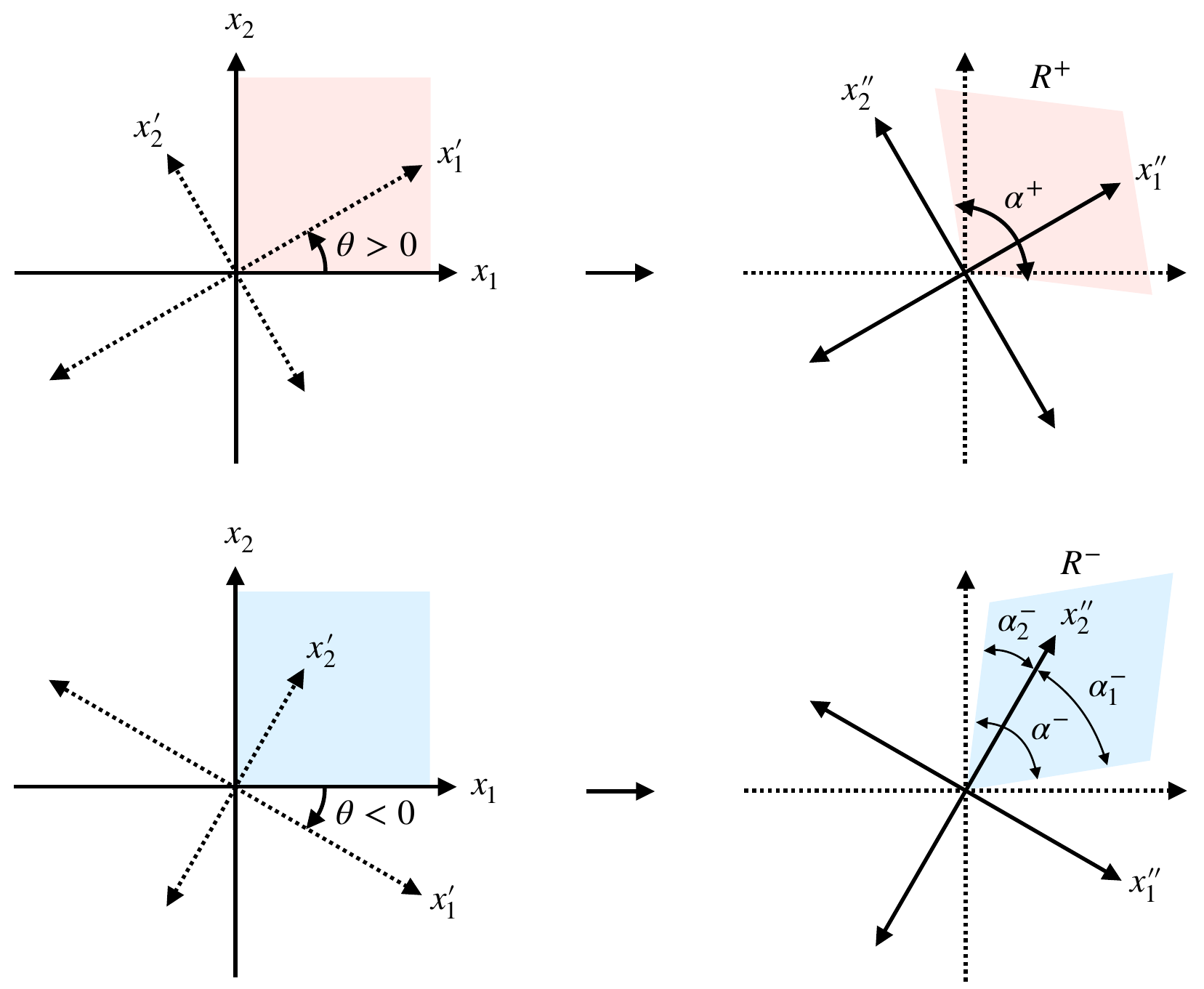}
\caption{\small
These plots represent the integration regions for $\vecbf{I}_1^+ \big|_{x_3 > x_4}$ (top) and $\vecbf{I}_1^- \big|_{x_3 > x_4}$ (bottom) before and after the coordinate transformations that are performed in the main text.
For the former, the original integration region $x_1 > 0$ and $x_2 > 0$ expands after transforming to the $x_1''$ and $x_2''$ coordinates, while it shrinks for the latter.
We call these deformed regions $R^\pm$.
}
\label{fig:rotation}
\end{center}
\end{figure}
%%%%%%%%%%%%%%%%
Then
\begin{align}
\vecbf{I}_1^\pm \big|_{x_3 > x_4}
&=
\frac{\Gamma_{4 - d + \nu}}{\Gamma_\nu} \int_{R^\pm} \dd x_1'' \dd x_2'' \int_{x_3 > x_4 > 0} \dd x_3 \dd x_4 \int_0^\infty \dd x_5~\delta \left( 1 - \sum_{i \in I} x_i \right) \frac{x_5^{- 1 + \nu}}{\sqrt{\lambda_1 \lambda_2}} \frac{{\cal U}^{4 - 3 d / 2 + \nu}}{({\cal F}^\pm)^{4 - d + \nu}}.
\end{align}
Let us have a closer look at $R^\pm$.
The rotation angle and the eigenvalues satisfy
\begin{align}
\sin 2 \theta
&=
\mp \frac{1}{2} \frac{1}{\lambda_1 - \lambda_2} \frac{x_5}{x_3 x_4 x_5},
~~~~~~
\cos 2 \theta
=
\frac{1}{4} \frac{1}{\lambda_1 - \lambda_2} \frac{x_4 - x_3}{x_3 x_4 x_5},
\end{align}
and
\begin{align}
\lambda_1 + \lambda_2
&=
\frac{1}{4} \frac{x_3 + x_4 + 2 x_5}{x_3 x_4 x_5},
~~~~~~
\lambda_1 \lambda_2
=
\frac{1}{16} \frac{x_3 x_4 + x_4 x_5 + x_5 x_3}{(x_3 x_4 x_5)^2}.
\end{align}
The rotation angle $\theta$ is positive for $\vecbf{I}_1^+ \big|_{x_3 > x_4}$, and negative for $\vecbf{I}_1^- \big|_{x_3 > x_4}$.
As illustrated in Fig.~\ref{fig:rotation}, the original integration range $x_1 > 0$ and $x_2 > 0$ translates into the deformed regions $R^\pm$ with angle $\alpha^\pm$.
Taking $\alpha^-$, the two angles are defined by
\begin{align}
\tan \alpha^-_1
&=
- \frac{1}{\tan \theta} \sqrt{\frac{\lambda_1}{\lambda_2}},
~~~~~~
\tan \alpha^-_2
=
- \tan \theta \sqrt{\frac{\lambda_1}{\lambda_2}}\,,
\end{align}
leading to
\begin{align}
&\tan \alpha^-
=
\frac{\tan \alpha^-_1 + \tan \alpha^-_2}{1 - \tan \alpha^-_1 \tan \alpha^-_2}
=
\frac{(x_3 x_4 + x_4 x_5 + x_5 x_3)^{1 / 2}}{x_5}.
\end{align}
and finally, noting that $0 < \alpha^- < \pi / 2$ and $\alpha^+ + \alpha^- = \pi$,
\begin{align}
\alpha^+
&=
\pi - \arctan \frac{(x_3 x_4 + x_4 x_5 + x_5 x_3)^{1 / 2}}{x_5}\,,
\\
\alpha^-
&=
\arctan \frac{(x_3 x_4 + x_4 x_5 + x_5 x_3)^{1 / 2}}{x_5}\,.
\end{align}
The same set of angles also appear for $\vecbf{I}_1^\pm \big|_{x_3 < x_4}$.
Since $x_1''$ and $x_2''$ appear in the integrand only through the combination $x_1''^2 + x_2''^2$, we may extend the integration region to the whole $x_1''$-$x_2''$ plane, and compensate it by multiplying with $\alpha^\pm / 2 \pi$
\begin{align}
\vecbf{I}_1^\pm
=
\frac{4 \Gamma_{4 - d + \nu}}{\Gamma_\nu} \int_{- \infty}^\infty &  \dd x_1'' \int_{- \infty}^\infty \dd x_2'' \int_0^\infty \dd x_3 \int_0^\infty \dd x_4 \int_0^\infty \dd x_5~\delta \left( 1 - \sum_{i \in I} x_i \right)
\nonumber \\
&
\times \frac{(x_3 x_4 + x_4 x_5 + x_5 x_3)^{4 - 3 d / 2 + \nu}}{(x_3 x_4 x_5)^{4 - d + \nu} [1 - (x_1''^2 + x_2''^2) - i 0 ]^{4 - d + \nu}} \, \frac{\alpha^\pm(x_3,x_4,x_5)}{2 \pi}.
\end{align}
The $x_1''$ and $x_2''$ integrations can be performed to give (assuming $1\notin I$ and $2\notin I$)
\begin{align}
\vecbf{I}_1^\pm
=
- \frac{2 \Gamma_{3 - d + \nu}}{\Gamma_\nu} &\int_0^\infty \dd x_3 \int_0^\infty \dd x_4 \int_0^\infty \dd x_5~\delta \left( 1 - \sum_{i \in I} x_i \right)
\nonumber \\
&
\times \frac{x_5^{- 1 + \nu} (x_3 x_4 + x_4 x_5 + x_5 x_3)^{7 / 2 - 3 d / 2 + \nu}}{(x_3 x_4 x_5)^{3 - d + \nu}} \, \alpha^\pm(x_3,x_4,x_5).
\end{align}
To simplify the argument of $\arctan$, we insert
\begin{align}\label{app-delta-x1}
1
&=
\int_0^\infty \dd x_6~
\delta\!\left( x_6 - \frac{(x_3 x_4 + x_4 x_5 + x_5 x_3)^{1 / 2}}{x_5} \right),
\end{align}
and use the delta function to resolve the integration over $x_3$.
In order to proceed, let us write the delta function in \eqref{app-delta-x1} as the following equivalent form
\begin{align}\label{app-delta-x2}
\delta\!\left( x_6 - \frac{(x_3 x_4 + x_4 x_5 + x_5 x_3)^{1 / 2}}{x_5} \right)
= \frac{2 x_5^2 x_6}{x_4 + x_5}\,
\delta\!\left(x_3 - \frac{x_5^2 x_6^2 - x_4 x_5}{x_4 + x_5} \right).
\end{align}
It is now straightforward to integrate out $x_3$.
Note that the parameter $x_3 > 0$ implies $x_4 < x_5  x_6^2$ from the RHS of \eqref{app-delta-x2}.
We arrive at
\begin{align}
\vecbf{I}_1^\pm
=
- \frac{2 \Gamma_{3 - d + \nu}}{\Gamma_\nu} & \int_0^\infty \dd x_5 \int_0^\infty \dd x_6 \int_0^{x_5 x_6^2} \dd x_4  ~\delta \left( 1 - \sum_{i \in I} x_i \right)
\nonumber \\
&
\times \frac{2 x_5^2 x_6}{x_4 + x_5} \frac{x_5^{\nu - 1} (x_5 x_6)^{7 - 3 d + 2 \nu}}{(x_3 x_4 x_5)^{3 - d + \nu}} \times
\begin{cases}
\pi - \arctan x_6\,, &\mathrm{for}~\vecbf{I}_1^+,
\\[0.37 em]
\arctan x_6\,, &\mathrm{for}~\vecbf{I}_1^-,
\end{cases}
\end{align}
with $x_3$ implicitly being a rational function of $x_4$, $x_5$, and $x_6$, {\it cf}.\,\eqref{app-delta-x2}.
We may take $I = \{ 5 \}$ to perform the $x_5$ integration, and then integrate over $x_4$ and $x_6$ to get
\begin{align}
\vecbf{I}_1^\pm
&=
- \frac{4 \Gamma_{3 - d + \nu} \Gamma_{d - 2  - \nu}^2}{\Gamma_\nu \Gamma_{2d - 4 - 2\nu}} \int_0^\infty \dd x_6\, x_6^{d - 2  - 2 \nu}
\nonumber\\[-0.3 em]
&\hspace{35mm}
\times  {}_2F_1 \left(d - 2  - \nu, d - 2  - \nu; 2d - 4  - 2 \nu; - x_6^2 \right) \times
\begin{cases}
\pi - \arctan x_6
\\
\arctan x_6
\end{cases}
\nonumber \\
&=
\frac{\Gamma_{3 - d + \nu} \Gamma_{d - 2  - \nu}^2}{\Gamma_\nu}
\left[
-  \frac{\pi\, \Gamma_{(d- 3) / 2}^2 \Gamma_{(d- 1) / 2 - \nu}}{\Gamma_{d - 2  - \nu}^2 \Gamma_{(3 d - 7) / 2 - \nu}} \pm \frac{2 \pi \csc(\pi (d / 2 - \nu))}{1 - d + 2 \nu} \frac{1}{\Gamma_{2d - 4  - 2 \nu}}
\right.
\nonumber \\
&~~~~
\times {}_3F_2 \big(\tfrac{d}{2} - \tfrac{1}{2}  - \nu, d - 2  - \nu, d - 2  - \nu; 2d - 4  - 2 \nu, \tfrac{1}{2} + \tfrac{d}{2} - \nu; 1 \big)
\nonumber \\
&~~~~
\left.
\mp \, \frac{2 \Gamma_{d/2 - 1 }^2 \Gamma_{d/2 - 1  - \nu}}{\Gamma_{d - 2  - \nu}^2 \Gamma_{3d/2 - 3  - \nu}} \, _4F_3 \big( \tfrac{1}{2}, 1, \tfrac{d}{2} - 1 , \tfrac{d}{2} - 1; \tfrac{3}{2}, \tfrac{3 d}{2} - 3  - \nu, 2 - \tfrac{d}{2} + \nu; 1 \big)
\right].
\end{align}
In particular, for $\nu = 1$, the integral evaluates to \eqref{app-res-sunrise-2linear}, i.e.\,$\left.\vecbf{I}_1^{+}\right|_{\nu = 1} = 2 \vecbf{I}_1^{-}\big|_{\nu = 1} = - (4\pi / 3) \times \Gamma_{(d-3)/2}^3\, \Gamma_{4-d} / \Gamma_{(3d-9)/2}$.

\paragraph{The second integral $\vecbf{I}_2^\pm$}

The second integral $\vecbf{I}_2^\pm$ has almost the same Feynman pa\-ra\-metri\-za\-tion as the first one because of their identical topological structure when it comes to the squared propagators.
To be explicit, performing a shift for $\vecbf{\ell}_2$ according to $\vecbf{\ell}_2\to-(\vecbf{\ell}_{12}-\vecbf{q})$, $\vecbf{I}_2^\pm$ becomes
\begin{align}
\vecbf{I}_2^\pm
&=
\int{\dd^d\ell_1 \dd^d\ell_2 \over \pi^{d}}\, {1 \over (\ell_1^z)\, (\mp\ell_{2}^z)}\, {(\vecbf{q}^2)^{3 - d + \nu} \over \vecbf{\ell}_1^2\, [\vecbf{\ell}_2^2]^\nu\, (\vecbf{\ell}_{12} {-} \vecbf{q})^2\, }.
\end{align}
Note that $\vecbf{I}_1^\pm$ and $\vecbf{I}_2^\pm$ are directly related for $\nu = 1$, $\vecbf{I}_1^\pm|_{\nu = 1} = \vecbf{I}_2^\mp|_{\nu = 1}$.
For generic values of $\nu$, we obtain the following parametric representation for $\vecbf{I}_2^\mp$
\begin{align}
\vecbf{I}_2^\mp
&=
\frac{\Gamma_{4 - d + \nu}}{\Gamma_\nu} 
\Bigg(\prod_{i=1}^{5} \int_0^\infty \dd x_i\Bigg)
\delta\!\left( 1 - \sum_{i \in I} x_i \right) x_4^{\nu - 1} \, \frac{{\cal U}^{4 - 3 d / 2 + \nu}}{({\cal F}^\pm)^{4 - d + \nu}},
\end{align}
with the same Symanzik polynomials as before.
Thus, the only modification from $\vecbf{I}_1^\pm$ to $\vecbf{I}_2^\mp$ is to replace the factor $x_5^{- 1 + \nu}$ by $x_4^{- 1 + \nu}$
\begin{align}
\vecbf{I}_2^\pm  =
- \frac{2 \Gamma_{3 - d + \nu}}{\Gamma_\nu} \int_0^\infty & \dd x_5 \int_0^\infty \dd x_6 \int_0^{x_5 x_6^2} \dd x_4  ~\delta \left( 1 - \sum_{i \in I} x_i \right)
\nonumber \\
&
\times \frac{2 x_5^2 x_6}{x_4 + x_5} \frac{x_4^{- 1 + \nu} (x_5 x_6)^{7 - 3 d + 2 \nu}}{(x_3 x_4 x_5)^{3 - d + \nu}} \times
\begin{cases}
\arctan x_6 & \text{for}~\vecbf{I}_2^{+},
\\[0.37 em]
\pi - \arctan x_6\,& \text{for}~\vecbf{I}_2^{-}.
\end{cases}
\end{align}
Again we take $I = \{ 5 \}$ to perform the $x_5$ integration, and then integrate over $x_4$ and $x_6$ to get
\begin{align}
\vecbf{I}_2^\pm
&=
- \frac{4 \Gamma_{d - 3} \Gamma_{3 - d + \nu} \Gamma_{- 2 + d - \nu}}{\Gamma_\nu \Gamma_{- 4 + 2 d - 2 \nu}} \int_0^\infty \dd x_6\, x_6^{d - 4}
\nonumber \\
&\hspace{30mm}
\times  {}_2F_1 \left(d - 3, d - 2  - \nu; 2d - 5  - \nu; - x_6^2 \right) \times
\begin{cases}
\arctan x_6
\\[0.37 em]
\pi - \arctan x_6
\end{cases}
\nonumber \\
&=
\frac{\Gamma_{d - 3} \Gamma_{3 - d + \nu} \Gamma_{d - 2  - \nu}}{\Gamma_\nu} \left[ - 2^{4 - d} \pi^{3/2} \frac{\Gamma_{(d-3)/2} \Gamma_{{(d - 1)/2}  - \nu}}{\Gamma_{{d/2} - 1} \Gamma_{d - 2  - \nu} \Gamma_{{(3d - 7)/2} - \nu}} \right.
\nonumber\\
&~~~~
\pm \frac{2 \pi \csc(\pi d / 2)}{3 - d} \frac{1}{\Gamma_{2d - 5  - \nu}} \, _3F_2 \big(\tfrac{d}{2} - \tfrac{3}{2}, d - 3 , d - 2 - \nu; \tfrac{d}{2} - \tfrac{1}{2} , 2d - 5 - \nu; 1 \big)
\nonumber\\
&~~~~
\left. \pm  \frac{2^{5 - d} \pi^{1/2}\,\Gamma_{{d/2} - 2 } \Gamma_{{d/2} - \nu}}{\Gamma_{(d-3)/2} \Gamma_{d - 2  - \nu} \Gamma_{{3d/2} - 3  - \nu}} \, _4F_3 \big( \tfrac{1}{2}, 1,  \tfrac{d}{2} - 1, \tfrac{d}{2} - \nu; \tfrac{3}{2}, 3 - \tfrac{d}{2}, \tfrac{3 d}{2} - 3  - \nu; 1 \big) \right].
\end{align}
In particular, for $\nu = 1$, the integral evaluates to $\vecbf{I}_2^{\pm}\big|_{\nu = 1} = \vecbf{I}_1^{\mp}\big|_{\nu = 1}$.

\paragraph{Nontrivial relations} We finally comment on nontrivial relations among the hypergeometric functions that we have found with this procedure.
We have not been able to find the following two relations in the literature: $\left.\vecbf{I}_1^+\right|_{\nu = 1} = 2 \left.\vecbf{I}_1^-\right|_{\nu = 1} = \left.\vecbf{I}_2^-\right|_{\nu = 1} = 2 \left.\vecbf{I}_2^+\right|_{\nu = 1} = - (4 \pi / 3) \Gamma_{(d - 3)/2}^3 \Gamma_{4 - d} / \Gamma_{(3d - 9) / 2}$, and $\vecbf{I}_1^+ = 2 \vecbf{I}_2^+$ for generic $\nu$.
To clean up the notation we set $d = 2 a$ in the following.
The former identity implies
\begin{align}
&
\, {}_4F_3 \big( \tfrac{1}{2}, 1, a - 1 , a - 1 ; \tfrac{3}{2}, 3 - a, 3a - 4; 1 \big)
-\frac{\sin(\pi a)}{6} \frac{\Gamma_{3 - a} \Gamma_{a - 3 / 2 }^3 \Gamma_{3a - 4 }}{\Gamma_{a - 1}^2 \Gamma_{3 a - 9/2}}\\
&=
\frac{2^{4a - 7 }}{\pi} \frac{\Gamma_{3 - a} \Gamma_{a - 3 / 2 }^2 \Gamma_{3a - 4 }}{\Gamma_{4a - 5}} \, {}_3F_2 \big(a - \tfrac{3}{2} , 2a - 3, 2a - 3; a - \tfrac{1}{2}, 4a - 6; 1 \big)\,.\nn
\end{align}
The latter is equivalent to
\begin{align}
&
\frac{2\csc(\pi a)\, \Gamma_{a - \nu}}{\Gamma_{3 - a}}
\, {}_4F_3 \big( \tfrac{1}{2}, 1,a - 1 , a - \nu; \tfrac{3}{2}, 3 - a, 3a - 3  - \nu; 1 \big)
\nonumber \\
&
- \frac{\csc(\pi(a - \nu))\, \Gamma_{a - 1}}{\Gamma_{2 - a + \nu}}
\, {}_4F_3 \big( \tfrac{1}{2}, 1, a - 1, a - 1; \tfrac{3}{2}, 2 - a + \nu, 3a - 3  - \nu; 1 \big)
\nonumber \\
&=
- \frac{2^{2a - 3} \pi^{1 / 2} \csc(\pi a)\, \csc(\pi (2 a - \nu))\, \Gamma_{a - 3 / 2}\,\Gamma_{3a - 3  - \nu}}{(3 - 2 a)\, \Gamma_{4a - 5  - \nu}\,\Gamma_{3 - 2 a + \nu}} \, \nonumber \\
&~~~~~~
\times \, {}_3F_2 \big(a - \tfrac{3}{2}, 2a - 3, 2a - 2 - \nu; a - \tfrac{1}{2}, 4a - 5  - \nu; 1 \big)
\nonumber \\
&~~~
+ \frac{2^{5 - 4 a + 2 \nu} \pi^{3 / 2} \csc(\pi (a - \nu))\, \csc(\pi (2 a - \nu))\, \Gamma_{3a - 3  - \nu}}{(1 - 2 a + 2 \nu)\, \Gamma_{a - 1}\,\Gamma_{2a - 3 / 2  - \nu}\, \Gamma_{3 - 2 a + \nu}}
\nonumber \\
&~~~~~~
\times \, {}_3F_2 \big(a - \tfrac{1}{2} - \nu, 2a - 2 - \nu, 2a - 2  - \nu; \tfrac{1}{2} + a - \nu,  4a - 4  - 2 \nu; 1 \big)
\nonumber \\
&~~~
+ \frac{1}{2} \frac{\Gamma_{a - 3 / 2}^2\, \Gamma_{a - 1 / 2  - \nu}\, \Gamma_{3a - 3  - \nu}}{\Gamma_{a - 1 } \Gamma_{3a - 7/2  - \nu}} \, .
\end{align}
Whereas we were able to numerically confirm these identities, we leave an analytic proof for future research.

%%%%%%%%%%%%%%%%%%%%%%%%%%%%%%%%%%%%%%%%%%%%%%%%%%
\section{Wick rotations}\label{sec:appB}
%%%%%%%%%%%%%%%%%%%%%%%%%%%%%%%%%%%%%%%%%%%%%%%%%%
By default, \texttt{pySecDec} and \texttt{FIESTA} define loop integrals in Minkowski space.
To compute a Euclidean loop integral with these programs, one has to transform it into its Minkowskian counterpart by a (reverse) Wick rotation.

To proceed, we define the scalar product of two vectors as
\begin{align}\label{}
k_\mathsf{E}\cdot \ell_\mathsf{E} \equiv k_\mathsf{E}^0 \ell_\mathsf{E}^0 + \sum\nolimits_{j=1}^{d-1} k_\mathsf{E}^j \ell_\mathsf{E}^j
\qquad\text{or}\qquad
k_\mathsf{M}\cdot \ell_\mathsf{M} \equiv k_\mathsf{M}^0 \ell_\mathsf{M}^0 - \sum\nolimits_{j=1}^{d-1} k_\mathsf{M}^j \ell_\mathsf{M}^j
\end{align}
in $d$-dimensional Euclidean or Minkowski space respectively.
We relate them through the so-called Wick rotation 
\begin{align}\label{app-wick-rorate}
k_\mathsf{M}^0 = i k_\mathsf{E}^{0} ~~\text{and}~~ k_\mathsf{M}^j = k_\mathsf{E}^{j}
\qquad\Longrightarrow\qquad 
k_\mathsf{M}\cdot \ell_\mathsf{M} = - k_\mathsf{E}\cdot \ell_\mathsf{E}\,.
\end{align}
% Peskin 6.48  %Henn p83
Using this transformation, we can translate any integral from Euclidean space into Minkowski space, or vice versa.

Let us consider the following 2-loop example
\begin{align}\label{}
S_\mathsf{E}^\pm  = \int{\dd^d \ell_1^\mathsf{E}\, \dd^d \ell_2^\mathsf{E} \over \pi^d} 
&{1 \over 
(\ell_1^\mathsf{E}\cdot u^\mathsf{E} - i0) 
(\pm\ell_2^\mathsf{E}\cdot u^\mathsf{E} - i0)
}
\\
\times&
{1 \over
[(\ell_1^\mathsf{E})^2 - i0)]\,
[(\ell_2^\mathsf{E})^2 - i0)]\,
[(\ell_1^\mathsf{E} {+} \ell_2^\mathsf{E} {-} q^\mathsf{E})^2 - i0)]
}\,,
\nn
\end{align}
with $q_\mathsf{E}\cdot u_\mathsf{E} =0$.
According to the Wick rotation defined in \eqref{app-wick-rorate}, its Minkowskian counterpart reads
\begin{align}\label{}
S_\mathsf{M}^\pm  = \int{\dd^d \ell_1^\mathsf{M}\, \dd^d \ell_2^\mathsf{M} \over i^2 \pi^d} 
&{1 \over 
(-\ell_1^\mathsf{M}\cdot u^\mathsf{M} - i0) 
(\mp\ell_2^\mathsf{M}\cdot u^\mathsf{M} - i0)
}
\\
\times&
{1 \over
[-(\ell_1^\mathsf{M})^2 - i0)]\,[-(\ell_2^\mathsf{M})^2 - i0)]\,
[-(\ell_1^\mathsf{M} {+} \ell_2^\mathsf{M} {-} q^\mathsf{M})^2 - i0)]
}\,,
\nn
\end{align}
with $q_\mathsf{M}\cdot u_\mathsf{M} = - q_\mathsf{E}\cdot u_\mathsf{E} = 0$.
To show their equivalence explicitly, let us write down their parametric representations:
\begin{align}
\label{app-feyn-rep-e}
S_\mathsf{E}^\pm  &= i^5 \big(e^{-\frac{i \pi }{4}}\big)^{2d} 
\int_0^\infty \dd x_1 \int_0^\infty \dd x_2 \int_0^\infty \dd x_3 \int_0^\infty \dd x_4 \int_0^\infty \dd x_5\,
\mathcal{U}^{-d/2} e^{-i \mathcal{F}_\mathsf{E}^\pm/\mathcal{U}},
\\[0.35 em]
\label{app-feyn-rep-m}
S_\mathsf{M}^\pm  &= i^5 \big(e^{-\frac{i \pi }{4}}\big)^{2d}
\int_0^\infty \dd x_1 \int_0^\infty \dd x_2 \int_0^\infty \dd x_3 \int_0^\infty \dd x_4 \int_0^\infty \dd x_5\,
\mathcal{U}^{-d/2} e^{-i \mathcal{F}_\mathsf{M}^\pm/\mathcal{U}},
\end{align}
with
\begin{align}\label{}
\mathcal{U} &= x_4 x_5 +  x_3 x_4 +  x_3 x_5,
\\
\mathcal{F}_\mathsf{E}^\pm &=
q_\mathsf{E}^2\, x_3  x_4  x_5 - {1\over 4}\,u_\mathsf{E}^2 \big( x_1^2 ( x_4 {+}  x_5) + x_2^2 ( x_3 {+}  x_5) \mp 2 x_1 x_2  x_5\big),
\\
\mathcal{F}_\mathsf{M}^\pm &= 
-q_\mathsf{M}^2\, x_3  x_4  x_5 + {1\over 4}\,u_\mathsf{M}^2 \big( x_1^2 ( x_4 {+}  x_5) + x_2^2 ( x_3 {+}  x_5) \mp 2 x_1 x_2  x_5\big).
\end{align}
It is clear that the two expressions in \eqref{app-feyn-rep-e} and \eqref{app-feyn-rep-m} are identical because of $u_\mathsf{M}^2 = - u_\mathsf{E}^2$ and $q_\mathsf{M}^2 = - q_\mathsf{E}^2$ according to \eqref{app-wick-rorate}.
As discussed previously, $\mathcal{F}_\mathsf{E}^{-}$ can be positive if we take an unphysical value of $u_\mathsf{E}$ such that $u_\mathsf{E}^2=-1$.

\clearpage
\bibliographystyle{JHEP}
\bibliography{ref}

\end{document}